\documentclass[10pt,journal,compsoc]{IEEEtran}

\usepackage{listings}
\usepackage{verbatim}
\usepackage{stfloats}
\usepackage{textcomp}
\usepackage{booktabs, tabularx}
\usepackage{rotating}
\usepackage{xcolor}
\usepackage{fancybox}
 \usepackage{multirow}

\usepackage{url}

\usepackage[caption=false,font=footnotesize]{subfig}
\usepackage{balance}

\usepackage{array}

\newcommand{\peter}[1]{\textcolor{blue}{{\it [Peter says: #1]}}}
\newcommand{\ahmedPof}[1]{\textcolor{red}{}}

\newcommand{\heng}[1]{\textcolor{red}{{\it [Heng says: #1]}}}

\usepackage{xspace}
\newcommand{\approach}{{\em Logram}\xspace}

\usepackage{array}
\newcolumntype{\$}{>{\global\let\currentrowstyle\relax}}
\newcolumntype{^}{>{\currentrowstyle}}

\ifCLASSOPTIONcompsoc
  \usepackage[nocompress]{cite}
\else
  \usepackage{cite}
\fi
\ifCLASSINFOpdf
\else
\fi
\begin{document}
%
\title{\emph{Logram}: Efficient Log Parsing Using $n$-Gram Dictionaries}

\author{Hetong~Dai,~\IEEEmembership{Student Member,~IEEE,}
   Heng~Li,~\IEEEmembership{Member,~IEEE,}
   Che-Shao Chen,~\IEEEmembership{Student Member,~IEEE,}

   Weiyi~Shang,~\IEEEmembership{Member,~IEEE,}
    Tse-Hsun (Peter) Chen,~\IEEEmembership{Member,~IEEE,}
 
 \IEEEcompsocitemizethanks{\IEEEcompsocthanksitem Department of Computer Science and Software Engineering, Concordia University, Montreal,
  Canada.\protect\\
  E-mail: (he\_da, c\_chesha, shang, peterc)@encs.concordia.ca}

 \IEEEcompsocitemizethanks{\IEEEcompsocthanksitem School of Computing, Queen's University, Kingston,
  Canada.\protect\\
  E-mail: hengli@cs.queensu.ca
 
  }

 
}


%




\IEEEtitleabstractindextext{%
\begin{abstract}
Software systems usually record important runtime information in their logs. Logs help practitioners understand system runtime behaviors and diagnose field failures.
As logs are usually very large in size, automated log analysis is needed to assist practitioners in their software operation and maintenance efforts.
Typically, the first step of automated log analysis is log parsing, i.e., converting unstructured raw logs into structured data.
However, log parsing is challenging, because logs are produced by static templates in the source code (i.e., logging statements) yet the templates are usually inaccessible when parsing logs.
Prior work proposed automated log parsing approaches that have achieved high accuracy.
However, as the volume of logs grows rapidly in the era of cloud computing, efficiency becomes a major concern in log parsing.
In this work, we propose an automated log parsing approach, \approach, which leverages $n$-gram dictionaries to achieve efficient log parsing.
We evaluated \approach on 16 public log datasets and compared \approach with five state-of-the-art log parsing approaches.
We found that \approach achieves higher parsing accuracy than the best existing approaches and also outperforms these approaches in efficiency (i.e., 1.8 to 5.1 times faster than the second-fastest approaches in terms of end-to-end parsing time ). 
Furthermore, we deployed \approach on \emph{Spark} and we found that \approach scales out efficiently with the number of \emph{Spark} nodes (e.g., with near-linear scalability for some logs) without sacrificing parsing accuracy.
In addition, we demonstrated that \approach can support effective online parsing of logs, achieving similar parsing results and efficiency to the offline mode.
\end{abstract}

}


%
\maketitle
\IEEEdisplaynontitleabstractindextext
\IEEEpeerreviewmaketitle

\section{Introduction}
\label{sec:intro}
Modern software systems usually record valuable runtime information (e.g., important events and variable values) in logs.
Logs play an important role for practitioners to understand the runtime behaviors of software systems and to diagnose system failures~\cite{DBLP:conf/icse/BarikDDF16, DBLP:conf/sigsoft/CitoLFG15}. 
However, since logs are often very large in size (e.g., tens or hundreds of gigabytes)~\cite{DBLP:conf/dsn/OlinerS07, DBLP:conf/fast/SchroederG07}, prior research has proposed automated approaches to analyze logs. 
These automated approaches help practitioners with various software maintenance 
and operation activities, such as anomaly detection~\cite{DBLP:conf/sosp/XuHFPJ09,DBLP:conf/icdm/XuHFPJ09,DBLP:conf/usenix/LouFYXL10,DBLP:conf/icdm/FuLWL09,DBLP:conf/icsm/JiangHHF08}, failure diagnosis~\cite{DBLP:conf/msr/FuLLDZX13, Automate62:online}, performance diagnosis and improvement~\cite{DBLP:conf/nsdi/NagarajKN12, DBLP:conf/osdi/ChowMFPW14}, and system comprehension~\cite{DBLP:conf/msr/FuLLDZX13, DBLP:conf/icse/ShangJHAHM13}.
Recently, the fast-emerging AIOps (\underline{A}rtificial \underline{I}ntelligence for IT \underline{Op}eration\underline{s}) solutions also depend heavily on automated analysis of operation logs~\cite{DBLP:conf/icse/DangLH19, DBLP:conf/sigsoft/LinHDZSXLLWYCZ18, DBLP:conf/sigsoft/HeLLZLZ18, DBLP:conf/icdcs/El-SayedZS17, DBLP:conf/osdi/HuangGLZD18}.

\begin{figure}[!t]
    \centering
    \includegraphics[width=.98\columnwidth]{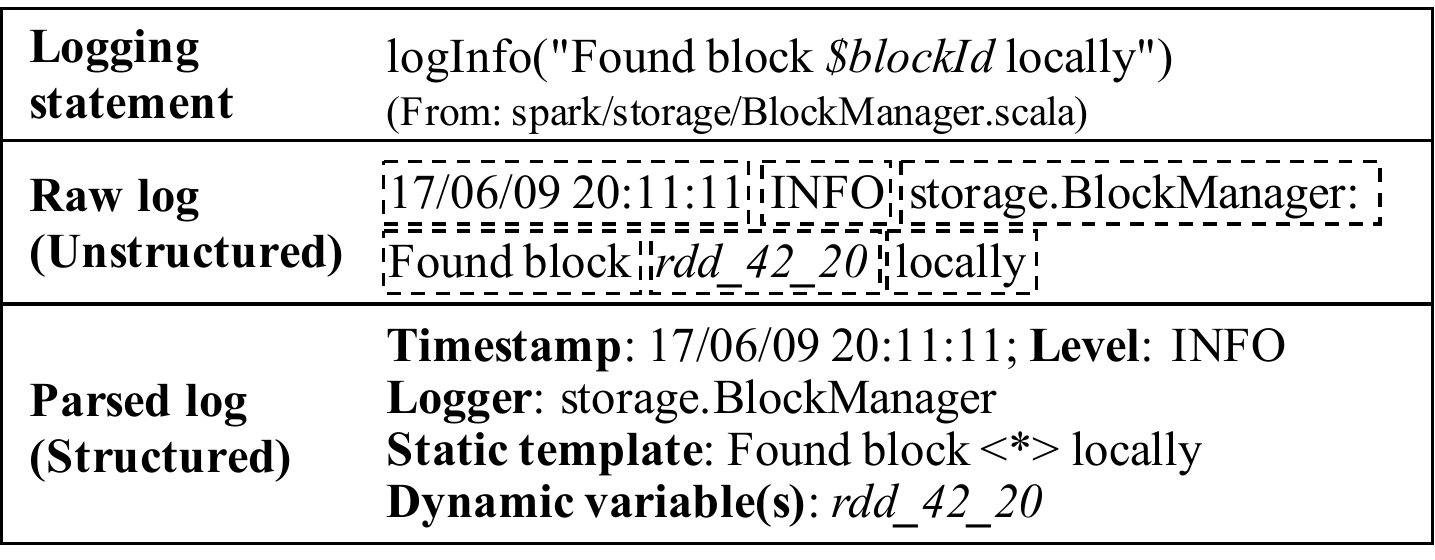}
    \caption{An illustrative example of parsing an unstructured log message into a structured format.}
    \label{fig:log-example}
\end{figure}

Logs are generated by logging statements in the source code.
As shown in Figure~\ref{fig:log-example}, a logging statement is composed of log level (i.e., {\em info}), static text (i.e., ``{\em Found block}'' and ``{\em locally}''), and dynamic variables (i.e., ``{\em \$blockId}''). 
During system runtime, the logging statement would generate raw log messages, which is a line of unstructured text that contains the static text and the values for the dynamic variables (e.g., ``{\em rdd\_42\_20}'') that are specified in the logging statement. The log message also contains information such as the timestamp (e.g., ``{\em 17/06/09 20:11:11}'') of when the event happened. 
In other words, logging statements define the templates for the log messages that are generated at runtime.
Automated log analysis usually has difficulties analyzing and processing the unstructured logs due to their dynamic nature~\cite{DBLP:conf/sosp/XuHFPJ09, DBLP:conf/msr/FuLLDZX13}. 
Instead, a log parsing step is needed to convert the unstructured logs into a structured format before the analysis.
The goal of log parsing is to extract the static template, dynamic variables, and the header information (i.e., timestamp, log level, and logger name) from a raw log message to a structured format.
Such structured information is then used as input for automated log analysis.
He et al.~\cite{7579781} found that the results of log parsing are critical to the success of log analysis tasks.

In practice, practitioners usually write \emph{ad hoc} log parsing scripts that depend heavily on specially-designed regular expressions~\cite{DBLP:conf/icse/ZhuHLHXZL19, DBLP:conf/icws/HeZZL17, DBLP:journals/smr/JiangHHF08} .
As modern software systems usually contain large numbers of log templates which are constantly evolving~\cite{shang2014exploratory, DBLP:conf/icse/YuanPZ12, DBLP:journals/ese/ChenJ17}, practitioners need to invest a  significant amount of efforts to develop and maintain such regular expressions. 
In order to ease the pain of developing and maintaining \emph{ad hoc} log parsing scripts, prior work proposed various approaches for automated log parsing~\cite{DBLP:conf/icse/ZhuHLHXZL19}. 
For example, \emph{Drain}~\cite{DBLP:conf/icws/HeZZL17} uses a fixed-depth tree to parse logs. Each layer of the tree defines a rule for grouping log messages (e.g., log message length, preceding tokens, and token similarity).
At the end, log messages with the same templates are clustered into the same groups.
Zhu et al.~\cite{DBLP:conf/icse/ZhuHLHXZL19} proposed a benchmark and thoroughly compared prior approaches for automated log parsing.


Despite the existence of prior log parsers, as the size of logs grows rapidly~\cite{DBLP:conf/wincom/LemouddenO15, DBLP:conf/sigsoft/CitoLFG15, DBLP:conf/icse/BarikDDF16} and the need for low-latency log analysis increases~\cite{DBLP:conf/icse/LiCHNF18, DBLP:conf/osdi/HuangGLZD18}, efficiency becomes an important concern for log parsing.
In this work, we propose \approach, an automated log parsing approach that leverages $n$-gram dictionaries to achieve efficient log parsing.
In short, \approach uses dictionaries to store the frequencies of $n$-grams in logs and leverage the $n$-gram dictionaries to extract the static templates and dynamic variables in logs. 
Our intuition is that frequent $n$-grams are more likely to represent the {\em static templates} while rare $n$-grams are more likely to be {\em dynamic variables}.
The $n$-gram dictionaries can be constructed and queried 
efficiently, i.e., with a complexity of $O(n)$ and $O(1)$, respectively. 

We evaluated \approach on 16 log datasets~\cite{DBLP:conf/icse/ZhuHLHXZL19} and compared \approach with five state-of-the-art log parsing approaches.
We found that \approach achieves higher accuracy compared with the best existing approaches, and that \approach outperforms these best existing approaches in efficiency, achieving a parsing speed that is 1.8 to 5.1 times faster than the second-fastest approaches.
Furthermore, as the $n$-gram dictionaries can be constructed in parallel and aggregated efficiently, we demonstrated that \approach can achieve high scalability when deployed on a multi-core environment (e.g., a \emph{Spark} cluster), without sacrificing any parsing accuracy.
Finally, we demonstrated that \approach can support effective online parsing, i.e., by updating the $n$-gram dictionaries continuously when new logs are added in a streaming manner.


In summary, the main contributions\footnote{The source code of our tool and the data used in our study are shared at \url{https://github.com/BlueLionLogram/Logram}} of our work 
include:
\begin{itemize}
    \item We present the detailed design of an innovative approach, \approach, for automated log parsing. \approach leverages $n$-gram dictionaries to achieve accurate and efficient log parsing.
    \item We compare the performance of \approach with other state-of-the-art log parsing approaches, based on an evaluation on 16 log datasets. The results show that \approach outperforms other state-of-the-art approaches in efficiency and achieves better accuracy than existing approaches.
    \item We deployed \approach on \emph{Spark} and we show that \approach scales out efficiently as the number of \emph{Spark} nodes increases (e.g., with near-linear scalability for some logs), without sacrificing paring accuracy.
    \item We demonstrate that \approach can effectively support online parsing, 
    achieving similar parsing results and efficiency compared to the offline mode.
    
    \item \approach automatically determines a threshold of $n$-gram occurrences to distinguish static and dynamic parts of log messages.
\end{itemize}

Our highly accurate, highly efficient, and highly scalable
\approach can benefit future research and practices that rely on automated log parsing for log analysis on large log data.
In addition, practitioners can leverage \approach in a log streaming environment to enable effective online log parsing for real-time log analysis.

\noindent\textbf{Paper organization.}
The paper is organized as follows. 
Section~\ref{sec:background} introduces the background of log parsing and $n$-grams.
Section~\ref{sec:related} surveys prior work related to log parsing. 
Section~\ref{sec:approach} presents a detailed description of our \approach approach.
Section~\ref{sec:evaluation} shows the results of evaluating \approach on 16 log datasets.
Section~\ref{sec:discussion} discusses the effectiveness of \approach for online log parsing.
Section~\ref{sec:threats} discusses the threats to the validity of our findings.
Finally, Section~\ref{sec:conclusion} concludes the paper.

\section{Background}
\label{sec:background}
In this section, we introduce the background of log parsing and $n$-grams that are used in our log parsing approach.

\subsection{Log Parsing} \label{sec:logparsing}

In general, the goal of log parsing is to extract the static template, dynamic variables, and the header information (i.e., timestamp, level, and logger) from a raw log message.
While the header information usually follows a fixed format that is easy to parse, extracting the templates and the dynamic variables is much more challenging, because 1) the static templates (i.e., logging statements) that generate logs are usually inaccessible~\cite{DBLP:conf/icse/ZhuHLHXZL19}, and 2) logs usually contain a large vocabulary of words~\cite{DBLP:journals/smr/JiangHHF08}.
Table~\ref{tab:log-exmaples} shows four simplified log messages with their header information removed.
These four log messages are actually produced from two static templates (i.e., ``{\em Found block $<$*$>$ locally}'' and ``{\em Dropping block $<$*$>$ from memory}''). 
These log messages also contain dynamic variables (i.e., ``{\em rdd\_42\_20}'' and ``{\em rdd\_42\_22}'') that vary across different log messages produced by the same template.
Log parsing aims to separate the static templates and the dynamic variables from such log messages. 

Traditionally, practitioners rely on \emph{ad hoc} regular expressions to parse the logs that they are interested in. 
For example, two regular expressions (e.g., ``{\em Found block [a-z0-9\_]+ locally}'' and ``{\em Dropping block [a-z0-9\_]+ from memory}'') could be used to parse the log messages shown in Table~\ref{tab:log-exmaples}.
Log processing \& management tools (e.g., \emph{Splunk}\footnote{https://www.splunk.com} and \emph{ELK} stack\footnote{https://www.elastic.co}) also enable users to define their own regular expressions to parse log data.
However, modern software systems usually contain large numbers (e.g., tens of thousands) of log templates which are constantly evolving~\cite{shang2014exploratory, DBLP:conf/icse/YuanPZ12, DBLP:journals/ese/ChenJ17, DBLP:journals/ese/LiSZH17, DBLP:conf/icse/ZhuHLHXZL19}.
Thus, practitioners need to invest a significant amount of efforts to develop and maintain such \emph{ad hoc} regular expressions.
Therefore, recent work proposed various approaches to automate the log parsing process~\cite{DBLP:conf/icse/ZhuHLHXZL19}.
In this work, we propose an automated log parsing approach that is highly accurate, highly efficient, highly scalable, and supports online parsing.

\begin{table}[!t]
\centering
\small
\caption{Simplified log messages for illustration purposes.}
\begin{tabular}{ll}
    \toprule
    1. & Found block \emph{rdd\_42\_20} locally \\
    2. & Found block \emph{rdd\_42\_22} locally \\
    3. & Dropping block \emph{rdd\_42\_20} from memory \\
    4. & Dropping block \emph{rdd\_42\_22} from memory \\
    \bottomrule
\end{tabular}
\label{tab:log-exmaples}
\end{table}

\subsection{$n$-grams} \label{sec:ngram}

An $n$-gram is a subsequence of length $n$ from an item sequence (e.g., text~\cite{cavnar1994n}, speech~\cite{DBLP:journals/taslp/SiuO00}, source code~\cite{DBLP:conf/wasa/NessaAWKQ08}, or genome sequences~\cite{DBLP:journals/cmpb/TomovicJK06}).
Taking the word sequence in the sentence: ``{\em The cow jumps over the moon}'' as an example, there are five 2-grams (i.e., bigrams): ``{\em The cow}'', ``{\em cow jumps}'', ``{\em jumps over}'', ``{\em over the}'', and ``{\em the moon}'', and four 3-grams (i.e., trigrams): ``{\em The cow jumps}'', ``{\em cow jumps over}'', ``{\em jumps over the}'', and ``{\em over the moon}''. 
$n$-grams have been successfully used to model natural language~\cite{cavnar1994n, DBLP:conf/naacl/LinH03, DBLP:journals/coling/BrownPdLM92, charniak1996statistical} and source code~\cite{hindle2012naturalness, Rahman:2019:NSR:3339505.3339511, Nguyen:2013:SSL:2491411.2491458}.
However, there exists no work that leverages $n$-grams to model log data.
In this work, we propose \approach that leverages $n$-grams to parse log data in an efficient manner.
Our intuition is that frequent $n$-grams are more likely to be static text while rare $n$-grams are more likely to be dynamic variables.

\approach extracts $n$-grams from the log data and store the frequencies of each $n$-gram in dictionaries (i.e., $n$-gram dictionaries).
Finding all the $n$-grams in a sequence (for a limited $n$ value) can be achieved efficiently by a single pass of the sequence (i.e., with a linear complexity)~\cite{DBLP:books/daglib/0022921}. For example, to get the 2-grams and 3-grams in the sentence ``{\em The cow jumps over the moon}'', an algorithm can move one word forward each time and get a 2-gram and a 3-gram starting from that word each time.
Besides, the nature of the $n$-gram dictionaries enables one to construct the dictionaries in parallel (e.g., by building separate dictionaries for different parts of logs in parallel and then aggregating the dictionaries).
Furthermore, the $n$-gram dictionaries can be updated online when more logs are added (e.g., in log streaming scenarios).
As a result, as shown in our experimental results, 
\approach is highly efficient, highly scalable, and supports online parsing.

\section{Related Work}
\label{sec:related}
In this section, we discuss prior work that proposed log parsing techniques and prior work that leveraged log parsing techniques in various software engineering tasks (e.g., anomaly detection).

\subsection{Prior Work on Log Parsing}
In general, existing log parsing approaches could be grouped under three categories: \emph{rule-based}, \emph{source code-based,} and \emph{data mining-based} approaches.

\noindent \textbf{Rule-based log parsing.} 
Traditionally, practitioners and researchers hand-craft heuristic rules (e.g., in the forms of regular expressions) to parse log data~\cite{vaarandi2006simple, damasio2002using, hansen1993automated}.
Modern log processing \& management tools usually provide support for users to specify customized rules to parse their log data~\cite{LogStash-Grok:online, Loggly-regex:online, Splunk-regex:online}.
Rule-based approaches require substantial human effort to design the rules and maintain the rules as log formats evolve~\cite{shang2014exploratory}.
Using standardized logging formats~\cite{W3C-Log-Format:online, ApacheNC79:online, IIS-Log-Format:online} can ease the pain of manually designing log parsing rules.
However, such standardized log formats have never been widely used in practice~\cite{DBLP:journals/smr/JiangHHF08}.

\noindent \textbf{Source code-based log parsing.} 
A log event is uniquely associated with a logging statement in the source code (see Section~\ref{sec:logparsing}).
Thus, prior studies proposed automated log parsing approaches that rely on the logging statements in the source code to derive log templates~\cite{DBLP:conf/sosp/XuHFPJ09, DBLP:conf/issre/NagappanWV09}.
Such approaches first use static program analysis to extract the log templates (i.e., from logging statements) in the source code.
Based on the log templates, these approaches automatically compose regular expressions to match log messages that are associated with each of the extracted log templates.
Following studies~\cite{xu2010experience, DBLP:conf/msr/SchipperAD19} applied ~\cite{DBLP:conf/sosp/XuHFPJ09} on production logs (e.g., Google's production logs) and achieved a very high accuracy.
However, source code is often not available for log parsing tasks, for example, when the log messages are produced by closed-source software or third-party libraries; not to mention the efforts for performing static analysis to extract log templates using different logging libraries or different programming languages.

\noindent \textbf{Data mining-based log parsing.} 
Other automated log parsing approaches do not require the source code, but instead, leverage various data mining techniques.
\emph{SLCT}~\cite{vaarandi2003data}, \emph{LogCluster}~\cite{DBLP:conf/msr/NagappanV10}, and \emph{LFA}~\cite{DBLP:conf/msr/NagappanV10} proposed approaches that automatically parse log messages by mining the frequent tokens in the log messages. These approaches count token frequencies and use a predefined threshold to identify the static components of log messages. The intuition is that if a log event occurs many times, then the static components will occur many times, whereas the unique values of the dynamic components will occur fewer times.
Prior work also formulated log parsing as a clustering problem and used various approaches to measure the similarity/distance between two log messages (e.g., \emph{LKE}~\cite{DBLP:conf/icdm/FuLWL09}, \emph{LogSig}~\cite{DBLP:conf/cikm/TangLP11}, \emph{LogMine}~\cite{Hamooni:2016:LFP:2983323.2983358}, \emph{SHISO}~\cite{DBLP:conf/IEEEscc/Mizutani13}, and \emph{LenMa}~\cite{DBLP:journals/corr/Shima16}). 
For example, \emph{LKE}~\cite{DBLP:conf/icdm/FuLWL09} clusters log messages into event groups based on the edit distance, weighted by token positions, between each pair of log messages. 

\emph{AEL}~\cite{DBLP:journals/smr/JiangHHF08} used heuristics based on domain knowledge to recognize dynamic components (e.g., tokens following the ``='' symbol) in log messages, then group log messages into the same event group if they have the same static and dynamic components.
\emph{Spell}~\cite{DBLP:conf/icdm/Du016} parses log messages based on a longest common subsequence algorithm, built on the observation that the longest common subsequence of two log messages are likely to be the static components. 
\emph{IPLoM}~\cite{DBLP:conf/kdd/MakanjuZM09} iteratively partitions log messages into finer groups, firstly by the number of tokens, then by the position of tokens, and lastly by the association between token pairs.
\emph{Drain}~\cite{DBLP:conf/icws/HeZZL17} uses a fixed-depth tree to represent the hierarchical relationship between log messages. Each layer of the tree defines a rule for grouping log messages (e.g., log message length, preceding tokens, and token similarity).
Zhu et al.~\cite{DBLP:conf/icse/ZhuHLHXZL19} evaluated the performance of such data mining-based parsers and they found that \emph{Drain}~\cite{DBLP:conf/icws/HeZZL17} achieved the best performance in terms of accuracy and efficiency.
Our $n$-gram-based log parser achieves a much faster parsing speed and a comparable parsing accuracy compared to \emph{Drain}.

\subsection{Applications of Log Parsing}
Log parsing is usually a prerequisite for various log analysis tasks, such as anomaly detection~\cite{DBLP:conf/sosp/XuHFPJ09,DBLP:conf/icdm/XuHFPJ09,DBLP:conf/usenix/LouFYXL10,DBLP:conf/icdm/FuLWL09,DBLP:conf/icsm/JiangHHF08}, failure diagnosis~\cite{DBLP:conf/msr/FuLLDZX13, Automate62:online}, performance diagnosis and improvement~\cite{DBLP:conf/nsdi/NagarajKN12, DBLP:conf/osdi/ChowMFPW14}, and system comprehension~\cite{DBLP:conf/msr/FuLLDZX13, DBLP:conf/icse/ShangJHAHM13}.
For example, Fu et al.~\cite{DBLP:conf/icdm/FuLWL09} first parse the raw log messages to extract log events. Based on the extracted event sequences, they then learn a Finite State Automaton (FSA) to represent the normal work flow, which is in turn used to detect anomalies in new log files.
Prior work~\cite{7579781} shows that the accuracy of log parsing is critical to the success of log analysis tasks.
Besides, as the size of log files grows fast~\cite{DBLP:conf/wincom/LemouddenO15, DBLP:conf/sigsoft/CitoLFG15, DBLP:conf/icse/BarikDDF16}, a highly efficient log parser is important to ensure that the log analysis tasks can be performed in a timely manner.
In this work, we propose a log parsing approach that is not only accurate but also efficient, which can benefit future log analysis research and practices.

\section{Approach}
\label{sec:approach}

In this section, we present our automated log parsing approach that is designed using $n$-gram dictionaries.

\subsection{Overview of \approach}

\begin{figure*}[t]
    \centering
    \includegraphics[width=.9\textwidth]{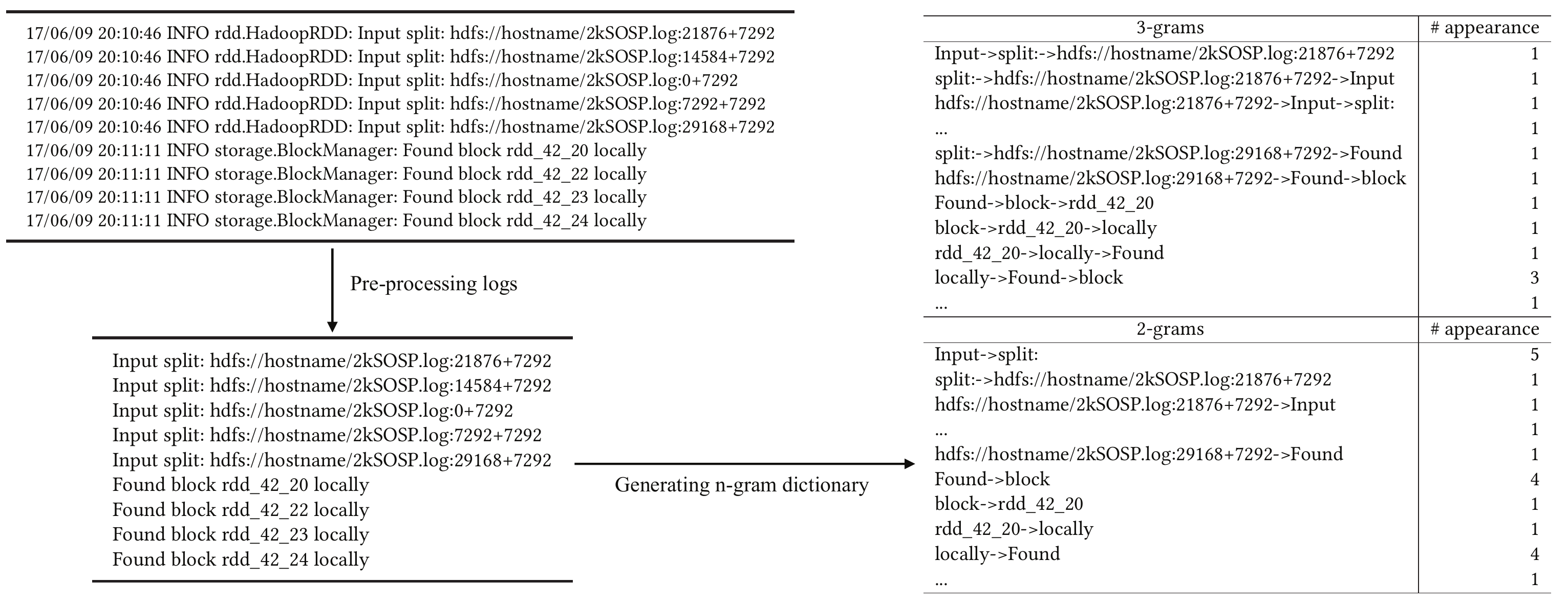}
        \vspace{-0.2cm}
    \caption{A running example of generating $n$-gram dictionary.}
    \label{fig:step1}
\end{figure*}

\begin{figure*}[t]
    \centering
    \includegraphics[width=.9\textwidth]{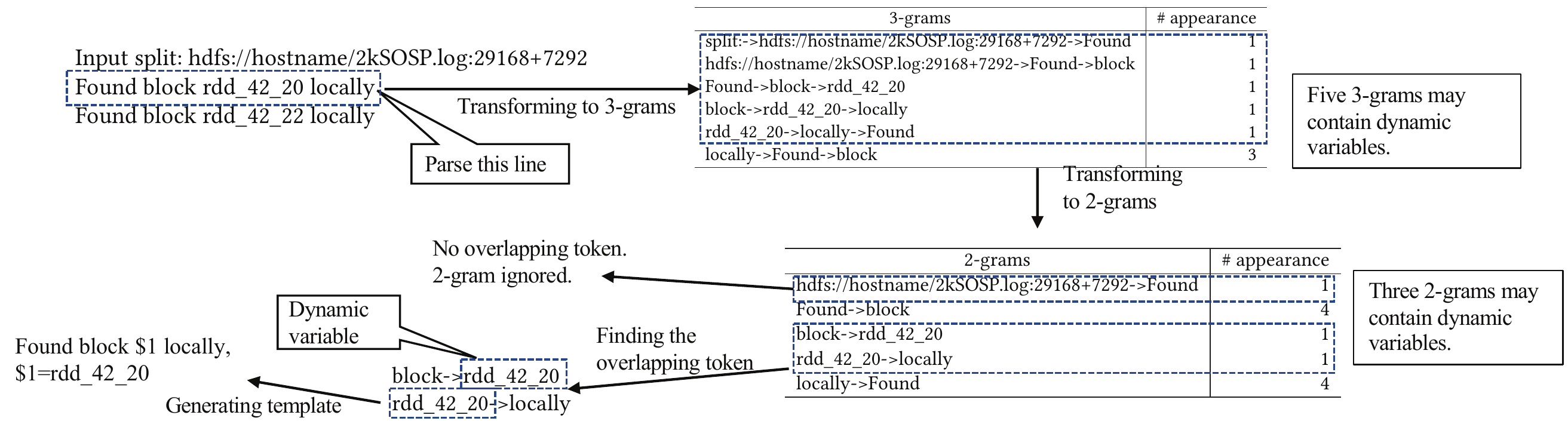}
    \vspace{-0.2cm}
    \caption{A running example of parsing one log message using the dictionary. The $n$-gram dictionary is abbreviated using ``...'' to avoid repetitive similar items.}
    \label{fig:step2}
\end{figure*}

Our approach consists of two steps: 1) generating $n$-gram dictionaries and 2) parsing log messages using $n$-gram dictionaries. In particular, the first step generates $n$-grams from log messages and calculate the number of appearances of each $n$-gram. In the second step, each log message is transformed into $n$-grams. By checking the number of appearance of each $n$-gram, we can automatically parse the log message into static text and dynamic variables. Figure~\ref{fig:step1} and \ref{fig:step2} show the overview of our approach with a running example.

\subsection{Generating an $n$-gram dictionary}

\subsubsection{Pre-processing logs} 

In this step, we extract a list of tokens (i.e., separated words) from each log message. First of all, we extract the content of a log message by using a pre-defined regular expression. For example, a log message often starts with the time stamp , the log level, and the logger name. Since these parts of logs often follow a common format in the same software system (specified by the configuration of logging libraries), we can directly parse and obtain these information. For example, a log message from the running example in Figure~\ref{fig:step1}, i.e., ``\emph{17/06/09 20:11:11 INFO storage.BlockManager: Found block rdd\_42\_24 locally}'', ``\emph{17/06/09 20:11:11}'' is automatically identified as time stamp, ``\emph{INFO}'' is identified as the log level and ``\emph{Storage.BlockManager:}'' is identified as the logger name; while the content of the log is ``\emph{Found block rdd\_42\_24 locally}''. After getting the content of each log message, we split the log message into tokens. The log message is split with white-space characters (e.g., \emph{space} and \emph{tab}). Finally, there exist common formats for some special dynamic information in logs, such as IP address and email address. 

In order to have a unbiased comparison with other existing log parsers in the \emph{LogPai} benchmark (cf. Section~\ref{sec:evaluation}), we leverage the openly defined regular expressions that are available from the \emph{LogPai} benchmark to identify such dynamic information.


\subsubsection{Generating an $n$-gram dictionary} 
\label{sec:generatingdic}

We use the log tokens extracted from each log message to create an $n$-gram dictionary.
Naively, for a log message with \emph{m} tokens, one may create an $n$-gram where $n\leq m$. However, when $m$ has the same value as $n$, the phrases with $n$-grams are exactly all log messages. Such a dictionary is not useful since almost all log messages have tokens that are generated from dynamic variables. On the other hand, a small value of $n$ may increase the chance that the text generated by a dynamic variable has multiple appearances. A prior study~\cite{he2018characterizing} finds that the repetitiveness of an $n$-gram in logs starts to become stable when $n\leq 3$. Therefore, in our approach, we generate the dictionary using phrases with two or three words (i.e., 2-grams and 3-grams). Naively, one may generate the dictionary by processing every single log message independently. However, such a naive approach has two limitations: 1) some log events may span across multiple lines and 2) the beginning and the ending tokens of a log message may not reside in the same number of $n$-grams like other tokens (c.f. our parsing step ``Identifying dynamically and statically generated tokens'' in Section~\ref{sec:identify}), leading to potential bias of the tokens being considered as dynamic variables. Therefore, at the beginning and the ending tokens of a log message, we also include the end of the prior log message and the beginning of the following log message, respectively, to create $n$-grams. For example, if our highest $n$ in the $n$-gram is 3, we would check two more tokens at the end of the prior log message and the beginning of the following log message. In addition, we calculate the number of occurrences of each $n$-gram in our dictionary.

As shown in a running example in Figure~\ref{fig:step1}, a dictionary from the nine lines of logs is generated consisting of 3-grams and 2-grams. Only one 3-grams, ``\emph{locally-$>$Found-$>$block}'', in the example have multiple appearance. Three 2-grams, ``\emph{Found-$>$block}'', ``\emph{Input-$>$split:}'' and ``\emph{locally-$>$Found}'', have four to five appearances. In particular, there exists $n$-grams, such as the 3-gram ``\emph{locally-$>$Found-$>$block}'', that are generated by combining the end and beginning of two log messages. Without such combination, tokens like ``\emph{input}'', ``\emph{Found}'' and ``\emph{locally}'' will have lower appearance in the dictionary.






\subsection{Parsing log messages using an $n$-gram dictionary}

In this step of our approach, we parse log messages using the dictionary that is generated from the last step. 

\subsubsection{Identifying $n$-grams that may contain dynamic variables}

 Similar to the last step, each log message is transformed into $n$-grams. For each $n$-gram from the log message, we check its number of appearances in the dictionary. If the number of occurrence of a $n$-gram is smaller than a automatically determined threshold (see Section~\ref{sec:threshold}), we consider that the $n$-gram may contain a token that is generated from dynamic variables. In order to scope down to identify the dynamically generated tokens, after collecting all low-appearing $n$-grams, we transform each of these $n$-grams into $n-1$-grams, and check the number of appearance of each $n-1$-gram. We recursively apply this step until we have a list of low-appearing 2-grams, where each of them may contain one or two tokens generated from dynamic variables. For our running example shown in Figure~\ref{fig:step2}, we first transform the log message into two 3-grams, while both only have one appearance in the dictionary. Hence, both 3-grams may contain dynamic variables. Afterwards, we transform the 3-grams into three 2-grams. One of the 2-grams (``\emph{Found-$>$block}'') has four appearances; while the other two 2-grams (``\emph{block-$>$rdd\_42\_20}'' and ``\emph{rdd\_42\_20-$>$locally}'') only have one appearance. Therefore, we keep the two 2-grams to identify the dynamic variables. 

\subsubsection{Identifying dynamically and statically generated tokens}
\label{sec:identify}

From the last step, we obtain a list of low-appearing 2-grams. However, not all tokens in the 2-grams are dynamic variables. There may be 2-grams that have only one dynamically generated token while the other token is static text. In such cases, the token from the dynamic variable must reside in two low-appearing 2-grams (i.e., one ends with the dynamic variable and one starts with the dynamic variable). For all other tokens, including the ones that are now selected in either this step or the last step, we consider them as generated from static text. 

However, a special case of this step is the beginning and ending tokens of each log message (c.f. the previous step ``Generating an $n$-gram dictionary'' in Section~\ref{sec:generatingdic}). Each of these tokens would only appear in smaller number of $n$-grams. For example, the first token of a log message would only appear in one 2-gram. If these tokens of the log message are from static text, they may be under-counted to be considered as potential dynamic variables. If these tokens of the log message are dynamically generated, they would never appear in two 2-grams to be identified as dynamic variable. To address this issue, for the beginning and ending tokens of each log message, we generate additional $n$-grams by considering the ending tokens from the prior log message; and for the ending tokens of each log message, we generate additional $n$-grams by considering the beginning tokens from the next log message. 

For our running example shown in Figure~\ref{fig:step2}, ``\emph{rdd\_42\_20}'' is generated from dynamic variable and it reside in two 2-grams (``\emph{block-$>$rdd\_42\_20}'' and ``\emph{rdd\_42\_20-$>$locally}''. Therefore, we can identify token ``\emph{rdd\_42\_20}'' as a dynamic variable, while ``\emph{block}'' and ``\emph{locally}'' are static text. On the other hand, since ``\emph{hdfs://hostname/2kSOSP.log}\\
\emph{:29168+7292-$>$Found}'' only appear without overlapping tokens with others, we ignore this 2-gram for identifying dynamic variables.

\subsubsection{Automatically determining the threshold of $n$-gram occurrences}
\label{sec:threshold}

The above identification of dynamically and statically generated tokens depends on a threshold of the occurrences of $n$-grams. In order to save practitioners' effort for manually searching the threshold, we use an automated approach to estimate the appropriate threshold. Our intuition is that most of the static $n$-grams would have more occurrences while the dynamic $n$-grams would have fewer occurrences. Therefore, there may exist a gap between the occurrences of the static $n$-grams and the occurrences of the dynamic $n$-grams, i.e., such a gap helps us identify a proper threshold automatically.

In particular, first, we measure the occurrences of each $n$-gram. Then, for each occurrence value, we calculate the number of $n$-grams that have the exact occurrence value. 
We use a two-dimensional coordinate to represent the occurrence values (i.e., the $X$ values) and the number of $n$-grams that have the exact occurrence values (i.e., the $Y$ values).
Then we use the \emph{loess} function~\cite{loessfun90:online} to smooth the $Y$ values and calculate the derivative of the $Y$ values against the $X$ values. 
After getting the derivatives, we use Ckmeans.1d.dp~\cite{Ckmeans138:online}, a one-dimensional clustering method, to find a break point to separate the derivatives into two groups, i.e., a group for static $n$-grams and a group for dynamic $n$-grams. The breaking point would be automatically determined as the threshold.

\subsubsection{Generating log templates} 

Finally, we generate log templates based on the tokens that are identified as dynamically or statically generated. We follow the same log template format as the \emph{LogPai} benchmark~\cite{DBLP:conf/icse/ZhuHLHXZL19}, in order to assist in further research. For our running example shown in Figure~\ref{fig:step2}, our approach parses the log message ``\emph{Found block rdd\_42\_20 locally}'' into ``\emph{Found block \$1 locally, \$1=rdd\_42\_20}''.

\section{Evaluation}
\label{sec:evaluation}
In this section, we present the evaluation of our approach. We evaluate our approach by parsing logs from the \emph{LogPai} benchmark~\cite{DBLP:conf/icse/ZhuHLHXZL19}. We compare \approach with five automated log parsing approaches, including \emph{Drain}~\cite{DBLP:conf/icws/HeZZL17}, \emph{Spell}~\cite{DBLP:conf/icdm/Du016}, \emph{AEL}~\cite{DBLP:journals/smr/JiangHHF08}, \emph{Lenma}~\cite{DBLP:journals/corr/Shima16} and \emph{IPLoM}~\cite{DBLP:conf/kdd/MakanjuZM09}
that are from prior research and all have been included in the \emph{LogPai} benchmark. We choose these five approaches since a prior study~\cite{DBLP:conf/icse/ZhuHLHXZL19} finds that these approaches have the highest accuracy and efficiency among all of the evaluated log parsers. In particular, we evaluate our approach on four aspects: 

\noindent \textbf{Accuracy.} The accuracy of a log parser measures whether it can correctly identify the static text and dynamic variables in log messages, in order to match log messages with the correct log events. A prior study~\cite{7579781} demonstrates the importance of high accuracy of log parsing, and low accuracy of log parsing can cause incorrect results (such as false positives) in log analysis. 

\noindent \textbf{Efficiency.} Large software systems often generate a large amount of logs during run time~\cite{Oliner:2012:ACL:2076450.2076466}. Since log parsing is typically the first step of analyzing logs, low efficiency in log parsing may introduce additional costs to practitioners when doing log analysis and cause delays to uncover important knowledge from logs.

\noindent \textbf{Ease of stabilisation.} Log parsers typically learn knowledge from existing logs in order to determine the static and dynamic components in a log message. The more logs seen, the better results a log parser can provide. It is desired for a log parser to have a stable result with learning knowledge from a small amount of existing logs, such that parsing the log can be done in a real-time manner without the need of updating knowledge while parsing logs. 

\noindent \textbf{Scalability.} Due to the large amounts of log data, one may consider leveraging parallel processing frameworks, such as \emph{Hadoop} and \emph{Spark}, to support the parsing of logs~\cite{he2017towards}. However, if the approach of a log parser is difficult to scale, it may not be adopted in practice. 


\subsection{Subject log data}

We use the data set from the \emph{LogPai} benchmark~\cite{DBLP:conf/icse/ZhuHLHXZL19}. The data sets and their descriptions are presented in Table~\ref{tab:subject}. The benchmark includes logs produced by both open source and industrial systems from various domains. These logs are typically used as subject data for prior log analysis research, such as system anomaly detection~\cite{7774521,Lin:2016:LCB:2889160.2889232}, system issue diagnosis~\cite{Du:2017:DAD:3133956.3134015} and system understanding~\cite{Oliner:2007:SSS:1251984.1253104}. To assist in automatically calculating accuracy on log parsing (c.f., Section\ref{sec:accuracy}), each data set in the benchmark includes a subset of 2,000 log messages that are already manually labeled with log event. Such manually labeled data is used in evaluating the accuracy of our log parser. For the other three aspects of the evaluation, we use the entire logs of each log data set.

\begin{table}[tbh]
    \caption{The subject log data used in our evaluation.}
    \small
    \label{tab:subject}
    \centering
    \begin{tabular}{c|c|r}
    \hline
    Platform & Description & Size \\
    \hline
    \hline
    Android & Android framework log & 183.37MB\\
    \hline
    Apache & Apache server error log & 4.90MB\\
    \hline
    BGL & Blue Gene/L supercomputer log & 708.76MB\\
    \hline
    Hadoop & Hadoop mapreduce job log & 48.61MB\\
    \hline
    HDFS & Hadoop distributed file system log & 1.47GB\\
    \hline
    HealthApp & Health app log & 22.44MB\\
    \hline
    HPC & High performance cluster log & 32.00MB\\
    \hline
    Linux & Linux system log & 2.25MB\\
    \hline
    Mac & Mac OS log & 16.09MB\\
    \hline
    OpenSSH & OpenSSH server log & 70.02MB\\
    \hline
    OpenStack & OpenStack software log & 58.61MB\\
    \hline
    Proxifier & Proxifier software log & 2.42MB\\
    \hline
    Spark & Spark job log & 2.75GB\\
    \hline
    Thunderbird & Thunderbird supercomputer log & 29.60GB\\
    \hline
    Windows & Windows event log & 26.09GB\\
    \hline
    Zookeeper & ZooKeeper service log & 9.95MB\\
    \hline
    \end{tabular}

\end{table}

\subsection{Accuracy}
\label{sec:accuracy}
In this subsection, we present the evaluation results on the accuracy of \approach. 

Prior approach by Zhu et al.~\cite{DBLP:conf/icse/ZhuHLHXZL19} defines an accuracy metric as the ratio of correctly parsed log messages over the total number of log messages. In order to calculate the parsing accuracy, a log event template is generated for each log message and log messages with the same template will be grouped together. If all the log messages that are grouped together indeed belong to the same log template, and all the log messages that indeed belong to this log template are in this group, the log messages are considered parsed correctly. However, the grouping accuracy has a limitation that it only determines whether the logs from the same events are grouped together; while the static text and dynamic variables in the logs may not be correctly identified. 

On the other hand, correctly identifying the static text and dynamic variables are indeed important for various log analysis. For example, Xu et al.~\cite{DBLP:conf/sosp/XuHFPJ09} consider the variation of the dynamic variables to detect run-time anomalies. Therefore, we manually check the parsing results of each log message and determine whether the static text and dynamic variables are correctly parsed, i.e., parsing accuracy. In other words, a log message is considered correctly parsed if and only if all its static text and dynamic variables are correctly identified. 


\begin{table}[tbh]
    \centering
        \caption{Accuracy of \approach compared with other log parsers. The results that are the highest among the parsers or higher than 0.9 are highlighted in bold. 
        }
        \small
    \label{tab:my_PA}


    \scalebox{0.9}{
    \begin{tabular}{c|c|c|c|c|c||c}
    \hline
    Name & \emph{Drain} & \emph{AEL} & \emph{Lenma} & \emph{Spell}  & \emph{IPLoM} & \emph{Logram} \\
    \hline
    \hline
    Android & \textbf{0.933} & 0.867 & \textbf{0.976} & \textbf{0.933}  & 0.716 & 0.848 \\
    \hline
    Apache & 0.693 & 0.693 & 0.693 & 0.693 & 0.693 &\textbf{0.699} \\
    \hline
    BGL &\textbf{ 0.822} & 0.818 & 0.577 & 0.639 & 0.792 & 0.740 \\
    \hline
    Hadoop & 0.545 & 0.539 & 0.535  & 0.192 & 0.373 & \textbf{0.965} \\
    \hline
    HDFS & \textbf{0.999} & \textbf{0.999} & \textbf{0.998} & \textbf{0.999}  & \textbf{0.998} & \textbf{0.981} \\
    \hline
    HealthApp & 0.609 & 0.615 & 0.141 & 0.602  & 0.651 &\textbf{0.969} \\
    \hline
    HPC & \textbf{0.929} & \textbf{0.990} & \textbf{0.915} & 0.694  & \textbf{0.979} &\textbf{0.959} \\
    \hline
    Linux & 0.250 & 0.241 & 0.251 & 0.131  & 0.235 &\textbf{0.460} \\
    \hline
    Mac & 0.515 & 0.579 & 0.551 & 0.434  & 0.503 &\textbf{0.666} \\
    \hline
    openSSH & 0.507 & 0.247 & 0.522 & 0.507 & 0.508 &\textbf{0.847} \\
    \hline
    Openstack & 0.538 & 0.718 & \textbf{0.759} & 0.592  & 0.697 &0.545 \\
    \hline
    Proxifier & \textbf{0.973} & \textbf{0.968} & \textbf{0.955} & 0.785  & \textbf{0.975} &\textbf{0.951} \\
    \hline
    Spark & \textbf{0.902} & \textbf{0.965} & \textbf{0.943} & 0.8645 & 0.883 &\textbf{0.903} \\
    \hline
    Thunderbird & 0.803 & 0.782 & \textbf{0.814} & 0.773  & 0.505 &0.761 \\
    \hline
    Windows & \textbf{0.983} & \textbf{0.983} & 0.277 & \textbf{0.978}  & 0.554 &\textbf{0.957} \\
    \hline
    Zookeeper & \textbf{0.962} & \textbf{0.922} & 0.842 & \textbf{0.955}  & \textbf{0.967} &\textbf{0.955} \\
    \hline
    \hline
    Average & 0.748 & 0.745 & 0.672 & 0.669 & 0.689 & \textbf{0.825}\\
    \hline
    \end{tabular}
    }
\end{table}

\subsubsection*{Results} \hfill

\noindent \textbf{\approach can achieve the best or over 0.9 accuracy in parsing 12 out of the 16 log datasets.}
Table~\ref{tab:my_PA} shows the accuracy on 16 datasets. Following the prior log-parsing benchmark~\cite{DBLP:conf/icse/ZhuHLHXZL19}, we highlight the accuracy results that are higher than 0.9, and highlight the highest accuracy in the same manner. We find that \approach has a higher or comparable accuracy compared to all other existing log parsers. On average, our approach has an accuracy of $0.825$, while the second highest accurate approach, i.e., \emph{Drain}, only has an average accuracy of $0.748$. In eight log data sets, \approach has an parsing accuracy higher than $0.9$ and in the four out of the rest eight datasets, \approach has the highest parsing accuracy among all parsers. Since \approach is designed based on processing every token in a log instead of comparing each line of logs with others, \approach exceeds other approaches in terms of parsing accuracy. 
In other words, even though prior approaches may often correctly group log messages together, the static text and dynamic variables from these log messages may not be correctly identified. Take the log parser, \emph{Spell} as an example. When parsing Hadoop logs the parsing accuracy is only 0.192. By manually checking the results, we find that some groups of logs share the string in host names that are generated from dynamic variables. For example, a log message ``\emph{Address change detected. Old: msra-sa-41/10.190.173.170:9000 New: msra-sa-41:9000}'' in the benchmark is parsed into a log template ``\emph{Address change detected. Old msra-sa-41/$<$*$>$ $<$*$>$ New msra-sa-41 $<$*$>$}''. We can see that the string ``\emph{msra-sa}'' in the host names is not correctly detected as dynamic variables. However, since all the log messages in this category have such string in the host names, even though \emph{Spell} cannot identify the dynamic variables, the log messages are still grouped together. 


Finally, by manually studying the incorrectly parsed log messages by \approach, we identify the following three reasons of incorrectly parsed log messages:
\begin{enumerate}
	\item \textbf{Mis-split tokens.} Some log messages contains special tokens such as \emph{$+$} and \emph{$\{$} to separate two tokens. In addition, sometimes static text and dynamic variables are printed into one single token without any separator (like white space). It is difficult for a token-based approach to address such cases.  
	
	\item \textbf{Pre-processing errors.} The pre-processing of common formats may introduce mistakes to the log parsing. For example, text in a common format (e.g., text of date) may be part of a long dynamic variable (a task id with its date as part of it). However, the pre-processing step extracts only the texts in the common format, causing the rest of the text in the dynamic variable parsed into a wrong format.
	
	\item \textbf{Frequently appearing dynamic variables.} Some dynamic variables contain contextual information of system environment and can appear frequently in logs. For example, in Apache logs, the path to a property file often appear in log messages such as: ``\emph{workerEnv.init() ok /etc/httpd/conf/workers2.properties}''. Although the path is a dynamic variable, in fact, the value of the dynamic variable never changes in the logs, preventing \approach from identifying it as a dynamic variable. On the other hand, such an issue is not challenging to address in practice. Practitioners can include such contextual information as part of pre-processing.
\end{enumerate}


\subsection{Efficiency}
\label{sec:efficiency}

To measure the efficiency of a log parser, similar to prior research~\cite{he2018directed,DBLP:conf/icse/ZhuHLHXZL19}, we record the elapsed time to finish the entire end-to-end parsing process on different log data with varying log sizes. We randomly extract data chunks of different sizes, i.e., 300KB, 1MB, 10MB, 100MB, 500MB and 1GB. Specifically, we choose to evaluate the efficiency from the Android, BGL, HDFS, Windows and Spark datasets, due to their proper sizes for such evaluation. From each log dataset, we randomly pick a point in the file and select a data chunk of the given size (e.g., 1MB or 10MB). We ensure that the size from the randomly picked point to the end of the file is not smaller than the given data size to be extracted. We measure the elapsed time for the log parsers on a desktop computer with Inter Core i5-2400 CPU \@ 3.10GHz CPU, 8GB memory and 7,200rpm SATA hard drive running Ubuntu 18.04.2 We compare \approach with three other parsers, i.e., \emph{Drain}, \emph{Spell} and \emph{AEL}, that all have high accuracy in our evaluation and more importantly, have the highest efficiency on log parsing based on the prior benchmark study~\cite{DBLP:conf/icse/ZhuHLHXZL19}.

\subsubsection*{Results} 

\begin{figure*}[tbh]
\centering
\subfloat[Android]{
    \includegraphics[trim=0 16 0 10,clip,width=.19\textwidth]{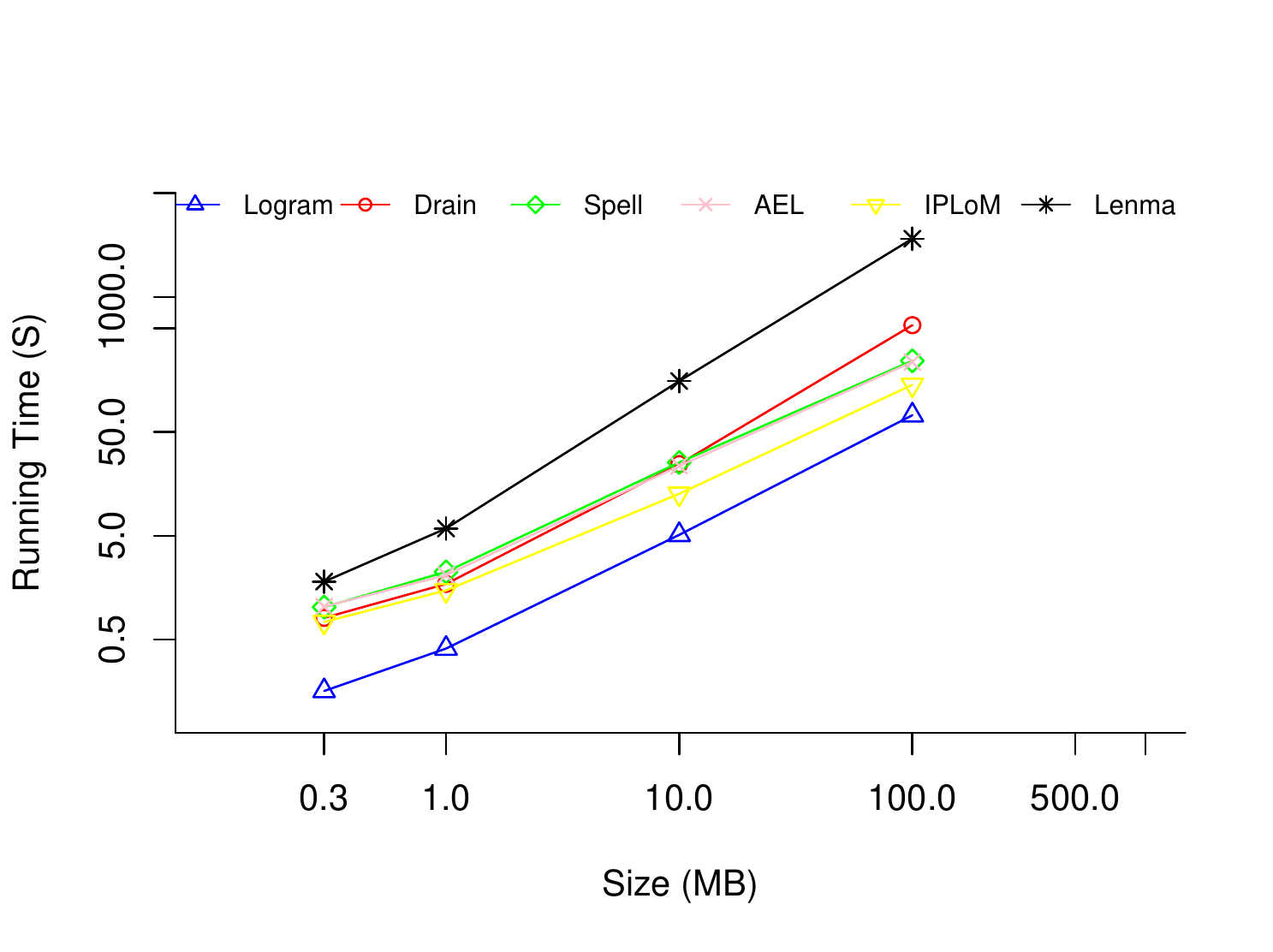}

}
\subfloat[BGL]{
    \includegraphics[trim=0 16 0 10,clip,width=.19\textwidth]{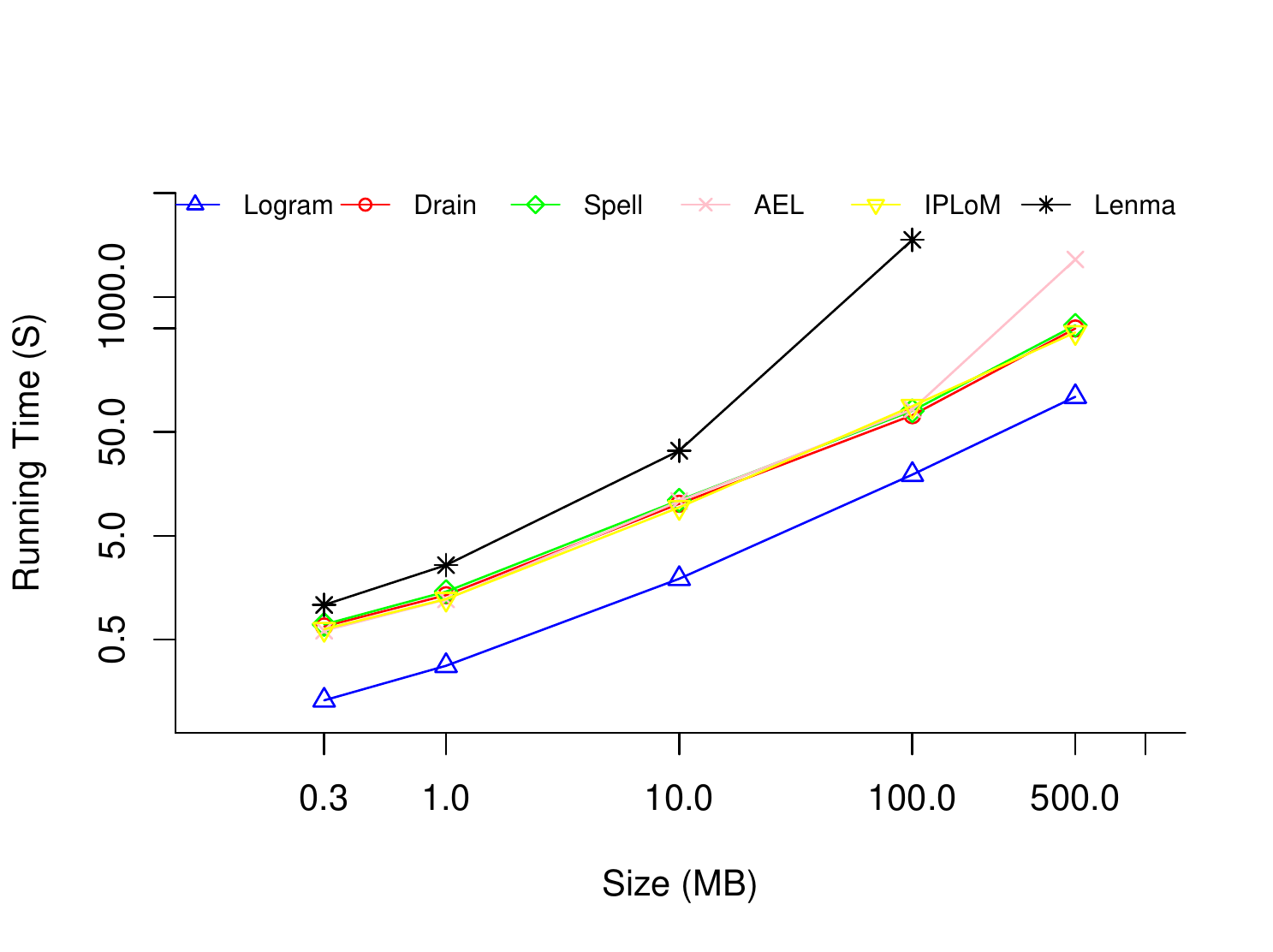}

}
\subfloat[HDFS]{
    \includegraphics[trim=0 16 0 10,clip,width=.19\textwidth]{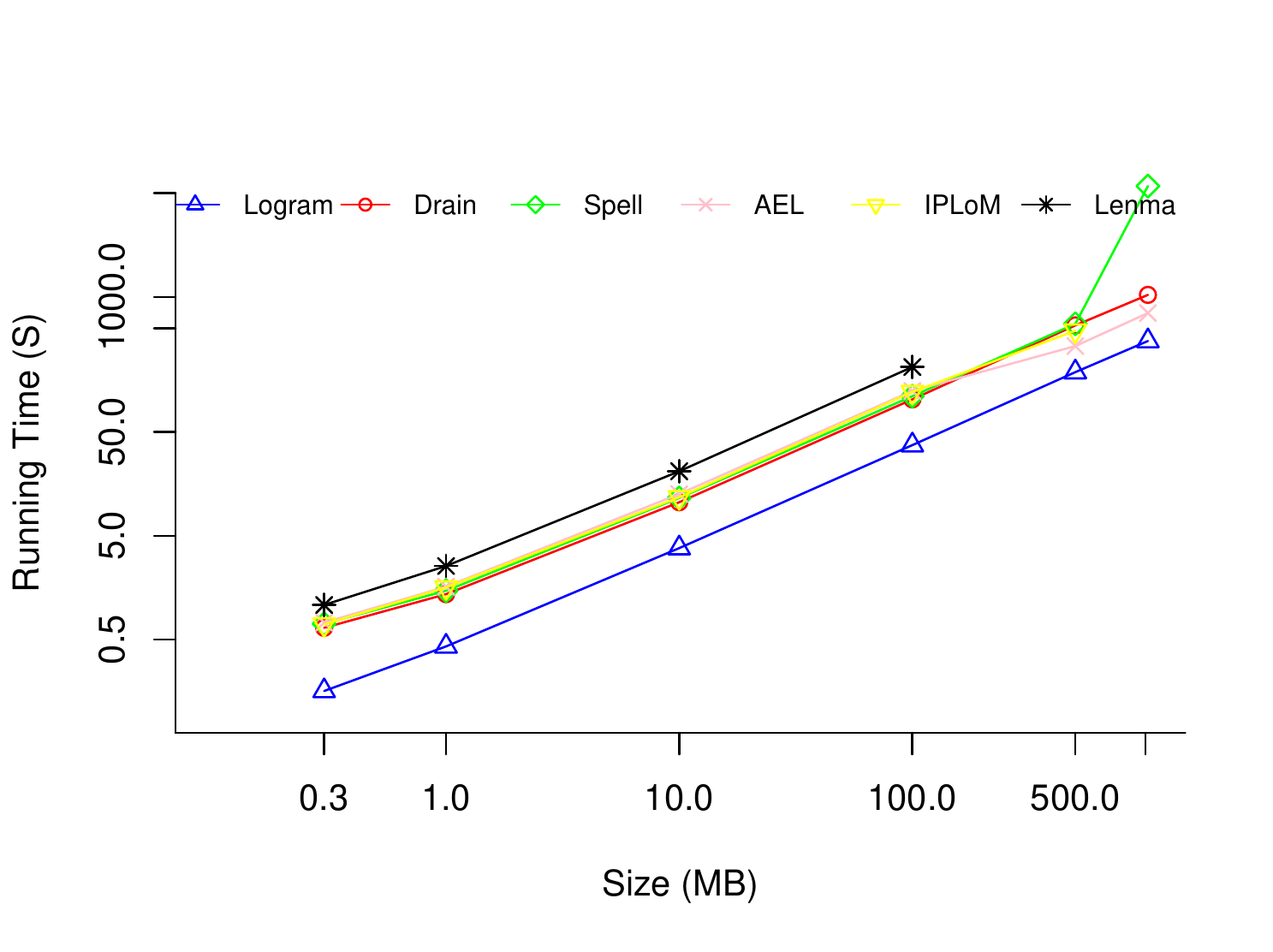}

}
\subfloat[Windows]{
    \includegraphics[trim=0 16 0 10,clip,width=.19\textwidth]{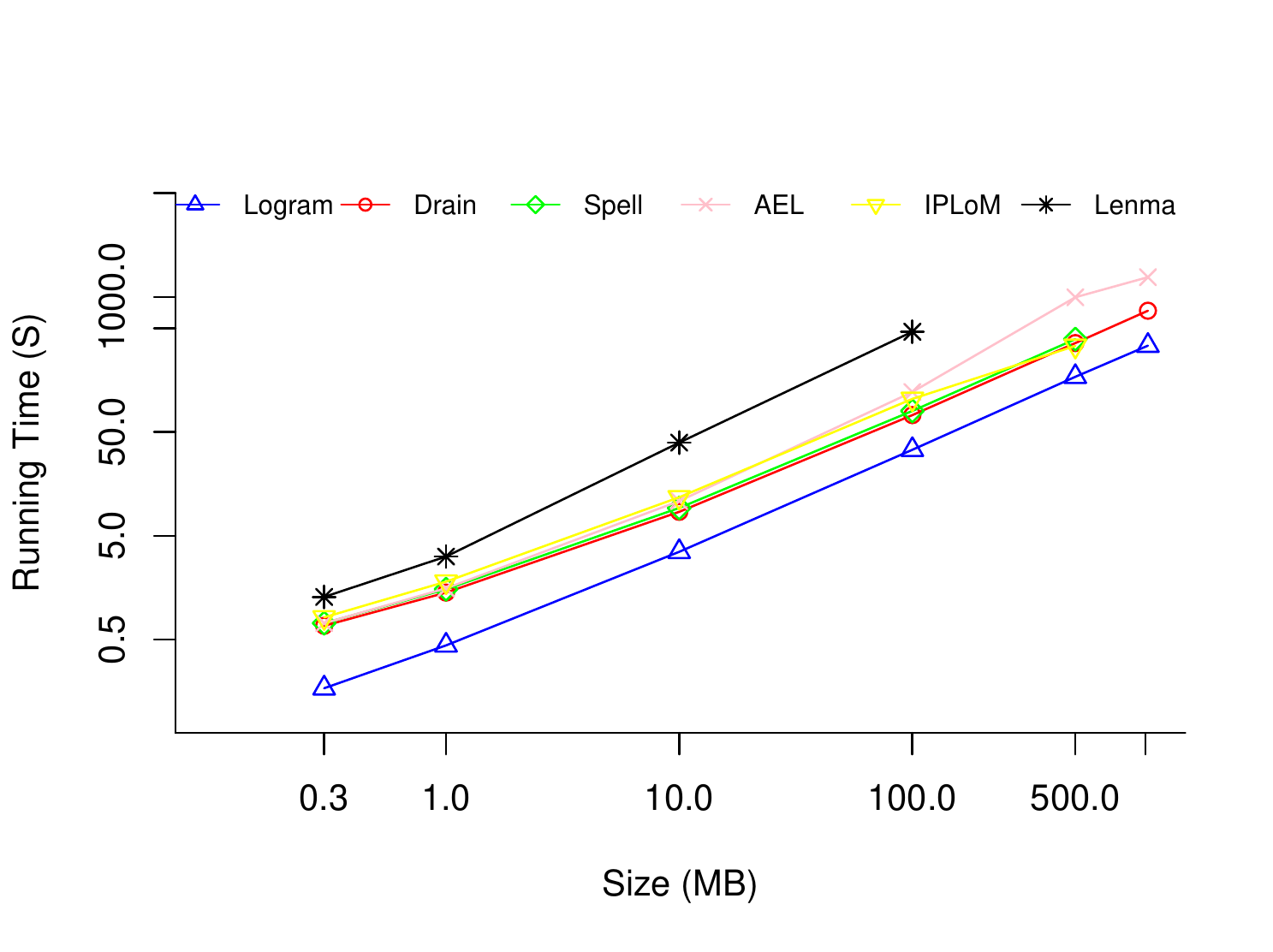}

}
\subfloat[Spark]{
    \includegraphics[trim=0 16 0 10,clip,width=.19\textwidth]{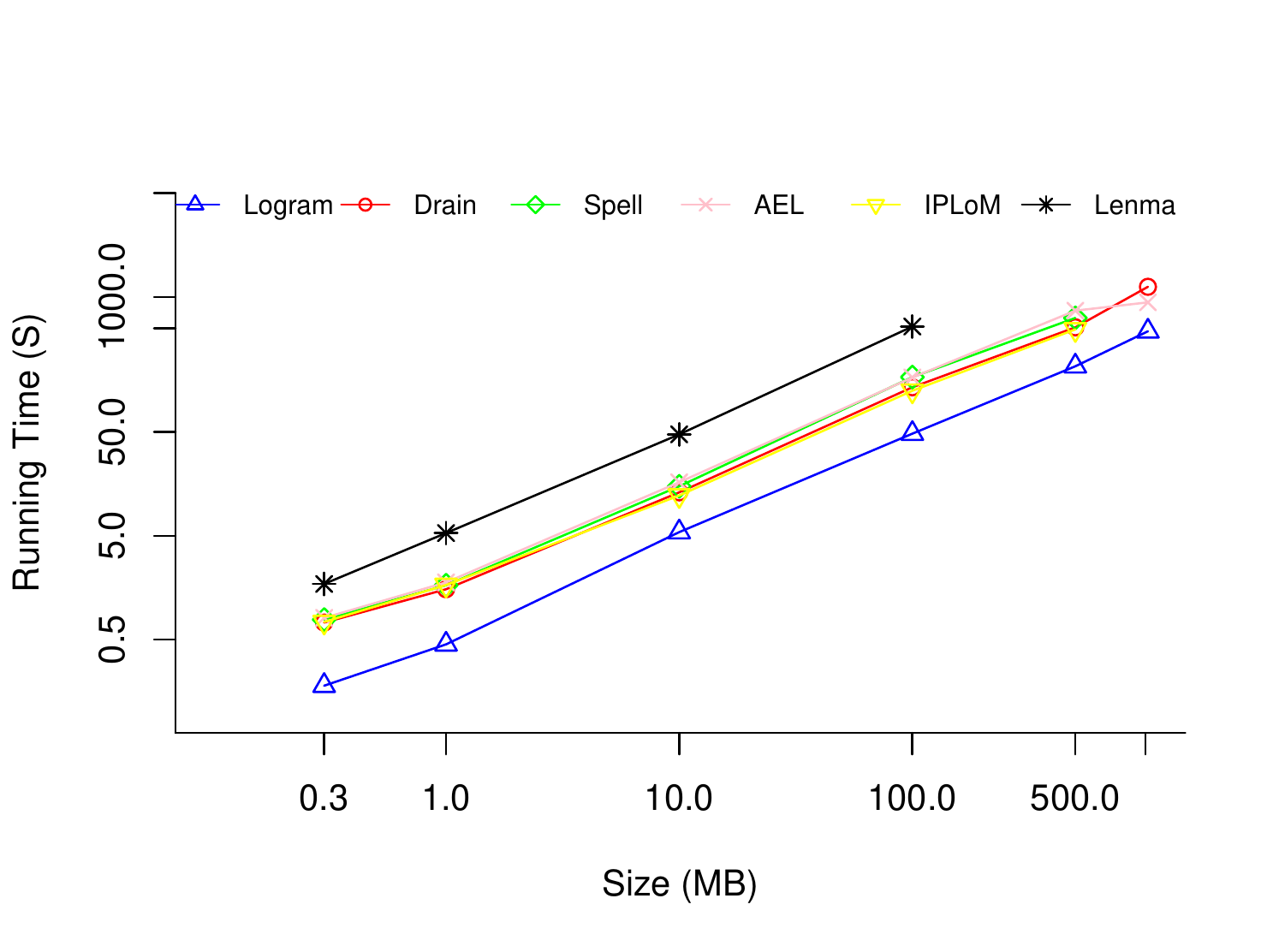}
}
\vspace{-0.1cm}
\caption{The elapsed time of parsing five different log data with various sizes. The x and y axes are in log scale.}
\label{figure:efficiency}
\end{figure*}

\textbf{\approach outperforms the fastest state-of-the-art log parsers in efficiency by 1.8 to 5.1 times.}
Figure~\ref{figure:efficiency} shows that time needed to parse five different log data with various sizes using \approach and five other log parsers. We note that the for \approach, the time to construct the dictionaries is already included in our end-to-end elapsed time, in order to fairly compare log parsing approaches. We find that \approach drastically outperform all other existing parsers. 
In particular, \approach is 1.8 to 5.1 times faster than the second fastest approaches when parsing the five log datasets along different sizes. 

\textbf{The efficiency of \approach is stable when increasing the sizes of logs.}
From our results, the efficiency of \approach is not observed to be negatively impacted when the size of logs increases.
For example, the running time only increases by a factor of 773 to 1039 when we increase the sizes of logs from 1 MB to 1GB (i.e., by a factor of 1,000).
\approach keeps a dictionary that is built from the $n$-grams in log messages. With larger size of logs, the size of dictionary may drastically increasing. However, our results indicate that up to the size of 1GB of the studied logs, the efficiency keeps stable. We consider the reason is that when paring a new log message using the dictionary, the look-up time for an $n$-gram in our dictionary is consistent, despite the size of the dictionary. Hence even with larger logs, the size of the dictionary do not drastically change. 

On the other hand, the efficiency of other log parsers may deteriorate with larger logs. In particular, \emph{Lenma} has the lowest efficiency among all studied log parsers. \emph{Lenma} cannot finish finish parsing any  500MB and 1GB log dataset within hours. In addition, \emph{Spell} would crash on Windows and Spark log files with 1G size due to memory issues. \emph{AEL} shows a lower efficiency when parsing large Windows and BGL logs. Finally, \emph{Drain}, although not as efficient as \approach, does not have a lower efficiency when parsing larger sizes of logs, which agrees with the finding in prior research on the log parsing benchmark~\cite{DBLP:conf/icse/ZhuHLHXZL19}.

\subsection{Ease of stabilisation}
\label{sec:stability}

We evaluate the ease of stabilisation by running \approach based on the dictionary from a subset of the logs. In other words, we would like to answer the following question: \\ \textbf{\emph{Can we generate a dictionary from a small size of log data and correctly parse the rest of the logs without updating the dictionary?}}

If so, in practice, one may choose to generate the dictionary with a small amount of logs without the need of always updating the dictionary while parsing logs, in order to achieve even higher efficiency and scalability. In particular, for each subject log data set, we first build the dictionary based on the first 5\% of the entire logs. Then we use the dictionary to parse the entire log data set. Due to the limitation of grouping accuracy found from the last subsection and the limitation of the high human effort needed to manually calculate the parsing accuracy, we do not calculate any accuracy for the parsing results. Instead, we automatically measure the agreement between the parsing result using the dictionary generated from the first 5\% lines of logs and the entire logs. For each log message, we only consider the two parsing results agree to each other if they are exactly the same. We then gradually increase the size of logs to build a dictionary by appending another 5\% of logs. We keep calculating the agreement until the agreement is 100\%. As running the experiments to repetitively parse the logs takes very long time, we evaluated the ease of stabilisation on 14 out of the 16 log datasets and we excluded the two largest log datasets (i.e., the Windows and Thunderbird logs).

\subsubsection*{Results} 

\begin{figure}[tbh]
\centering
\subfloat[Android]{
    \includegraphics[trim=0 10 0 0,clip,width=.32\columnwidth]{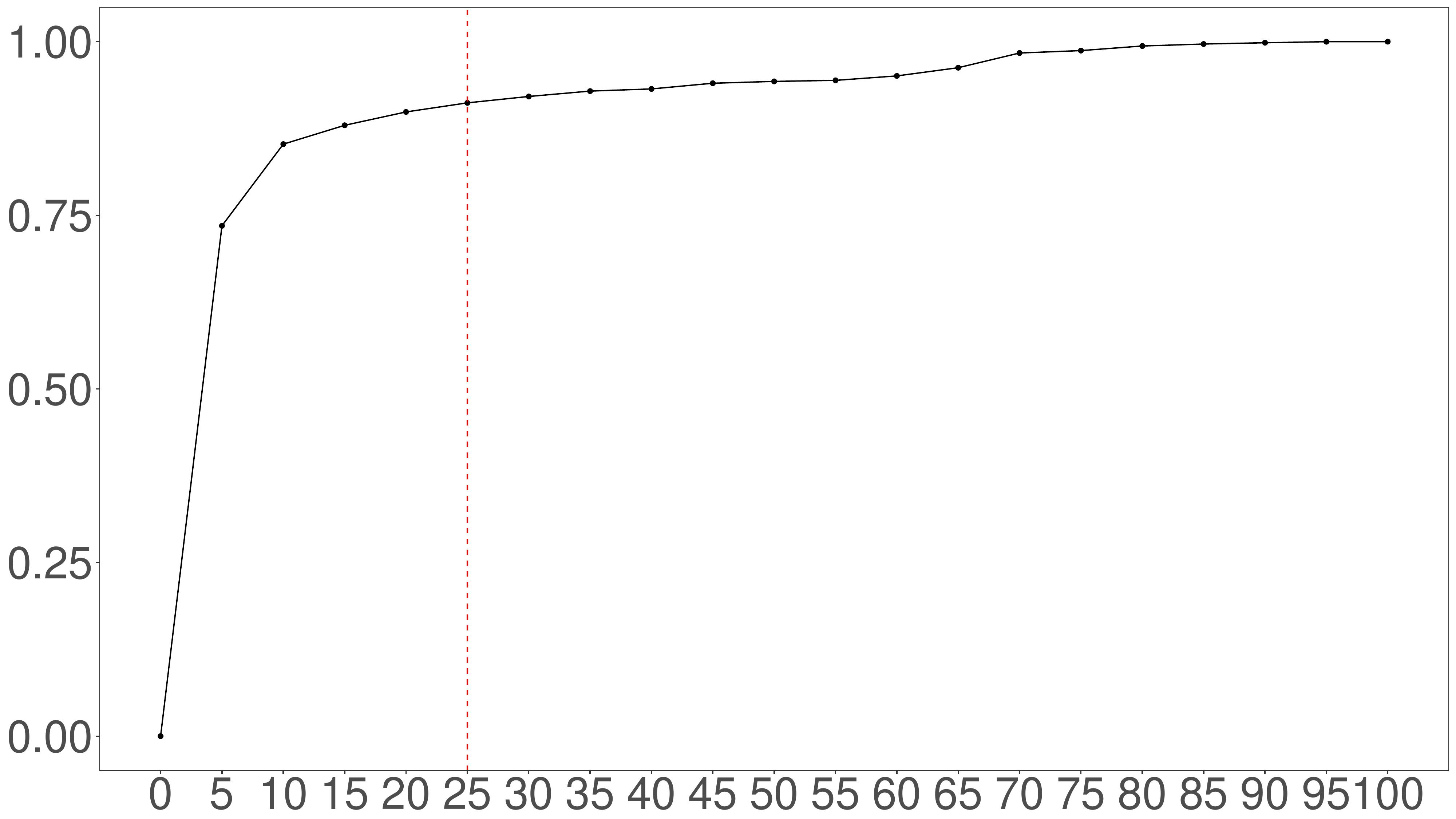}
}
\subfloat[Apache]{
    \includegraphics[trim=0 10 0 0,clip,width=.32\columnwidth]{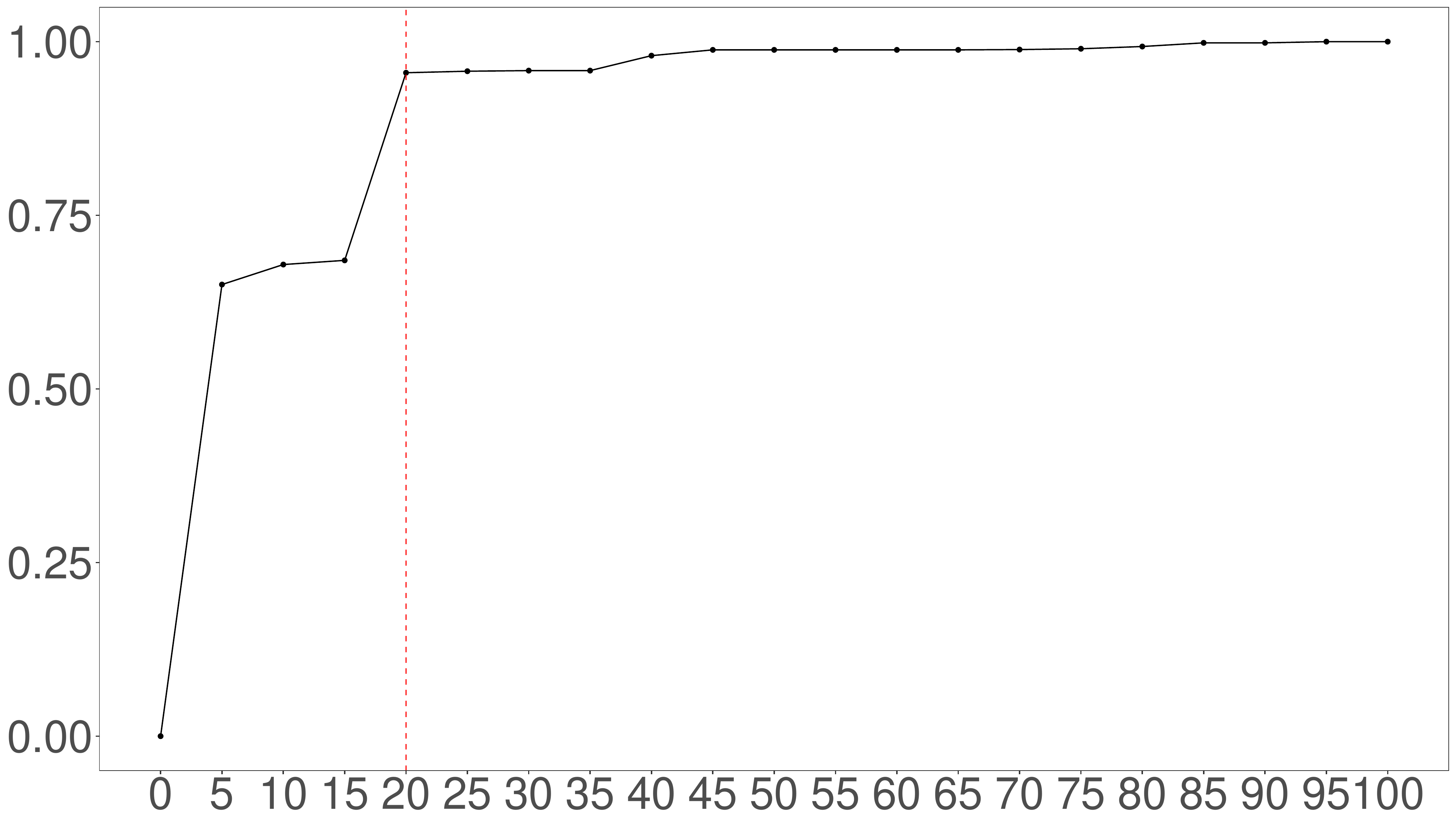}
}
\subfloat[BGL]{
    \includegraphics[trim=0 10 0 0,clip,width=.32\columnwidth]{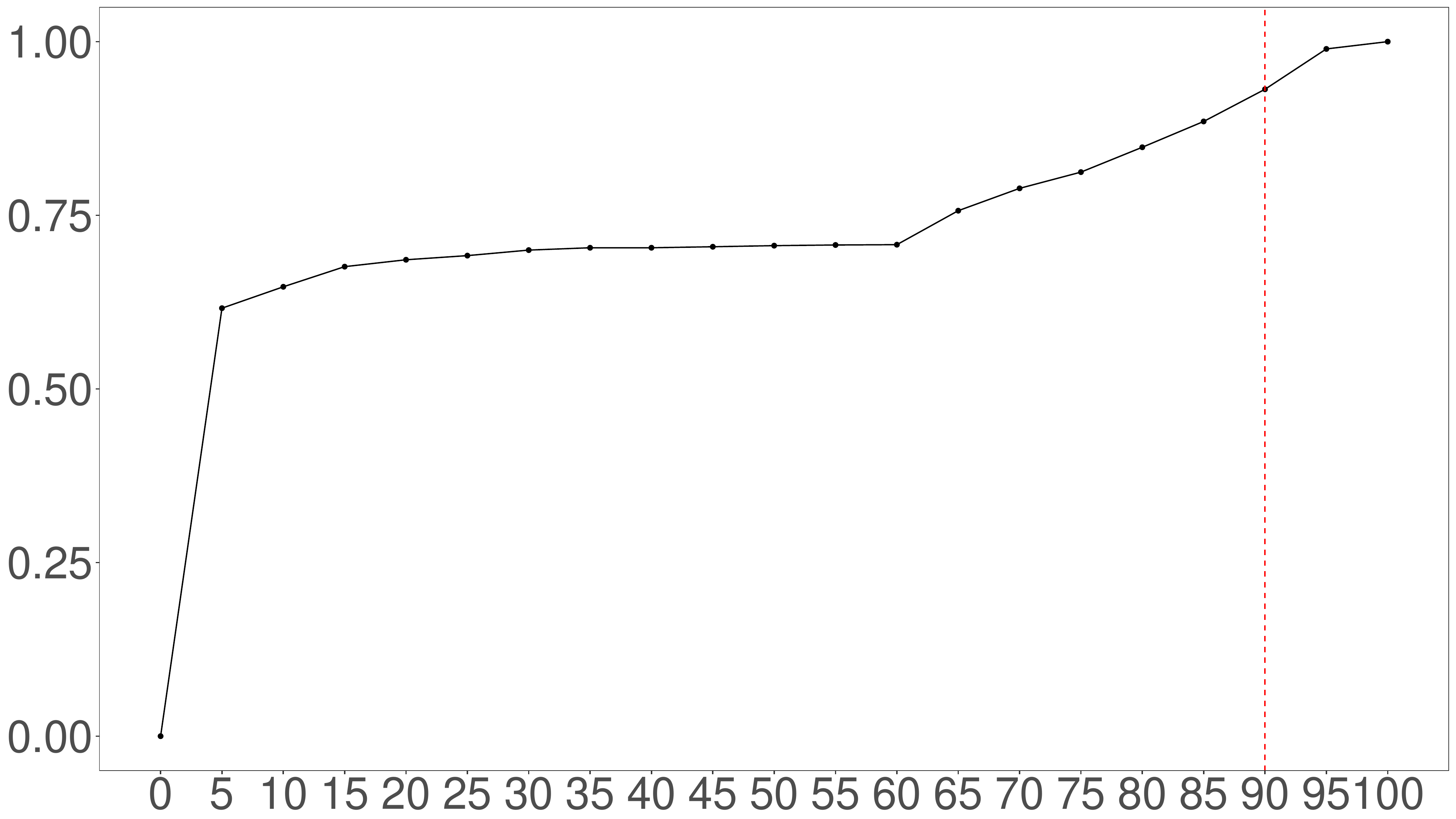}
}\\
\subfloat[Hadoop]{
    \includegraphics[trim=0 10 0 0,clip,width=.32\columnwidth]{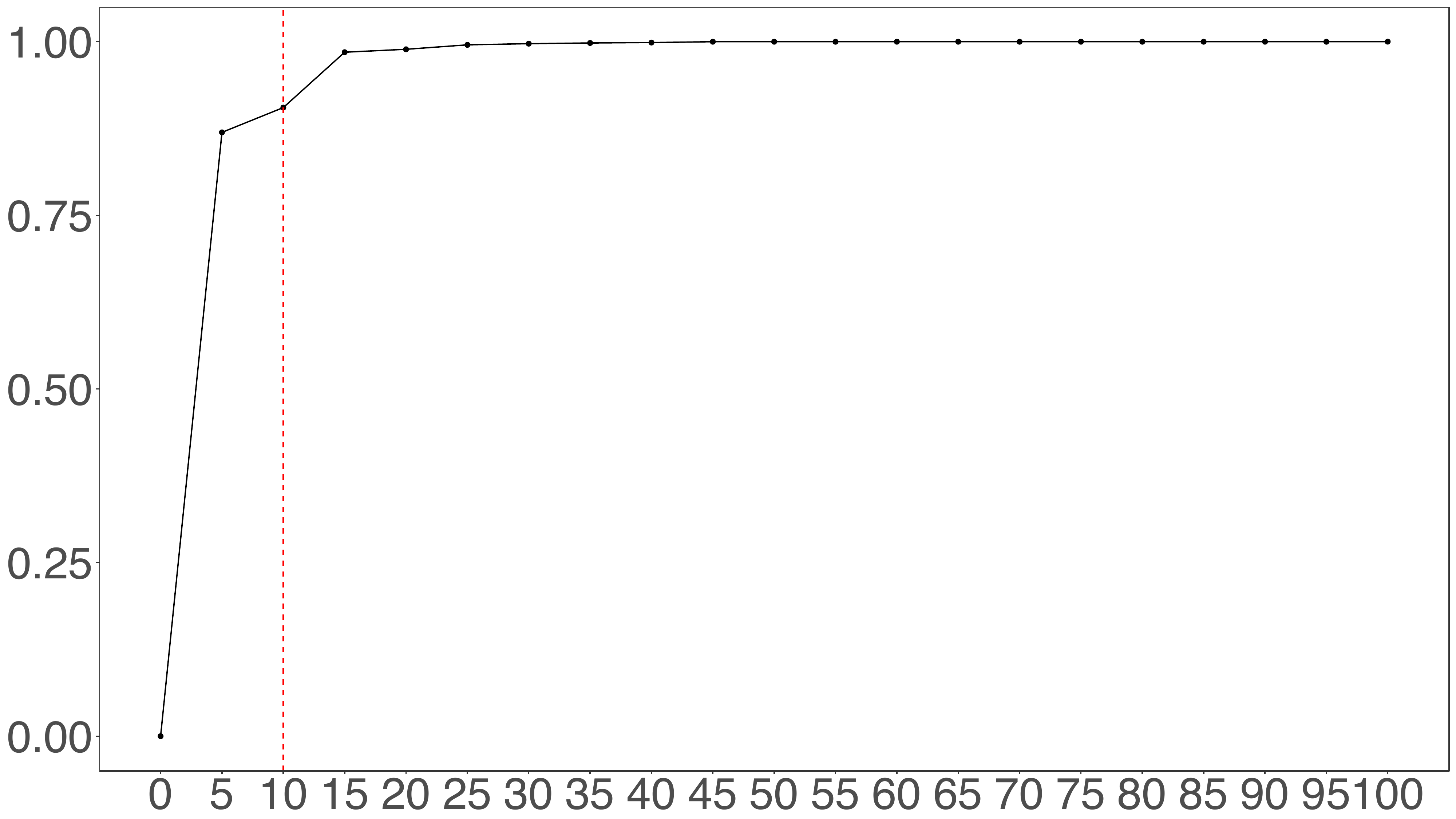}
}
\subfloat[HDFS]{
    \includegraphics[trim=0 10 0 0,clip,width=.32\columnwidth]{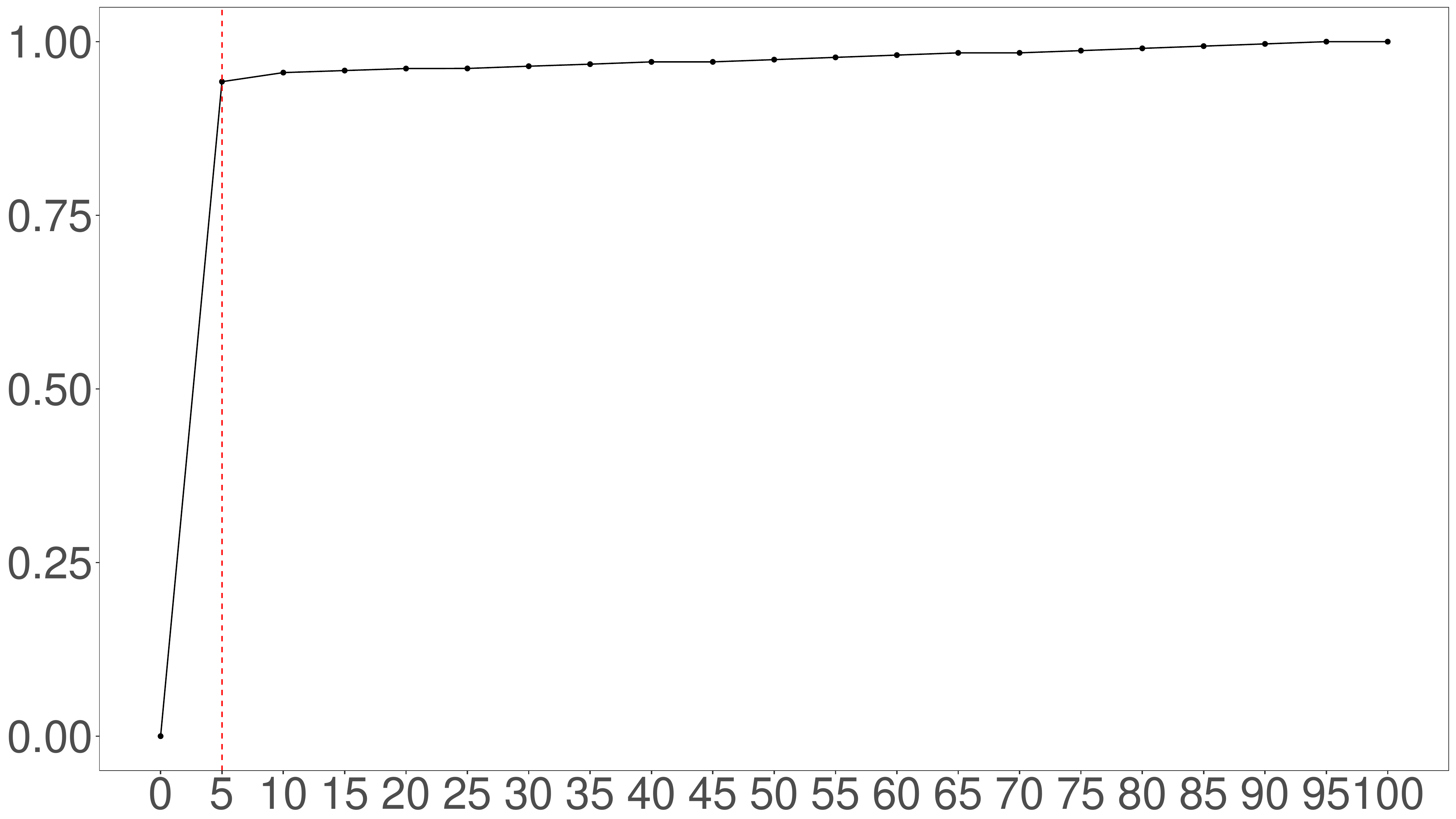}
}
\subfloat[HealthApp]{
    \includegraphics[trim=0 10 0 0,clip,width=.32\columnwidth]{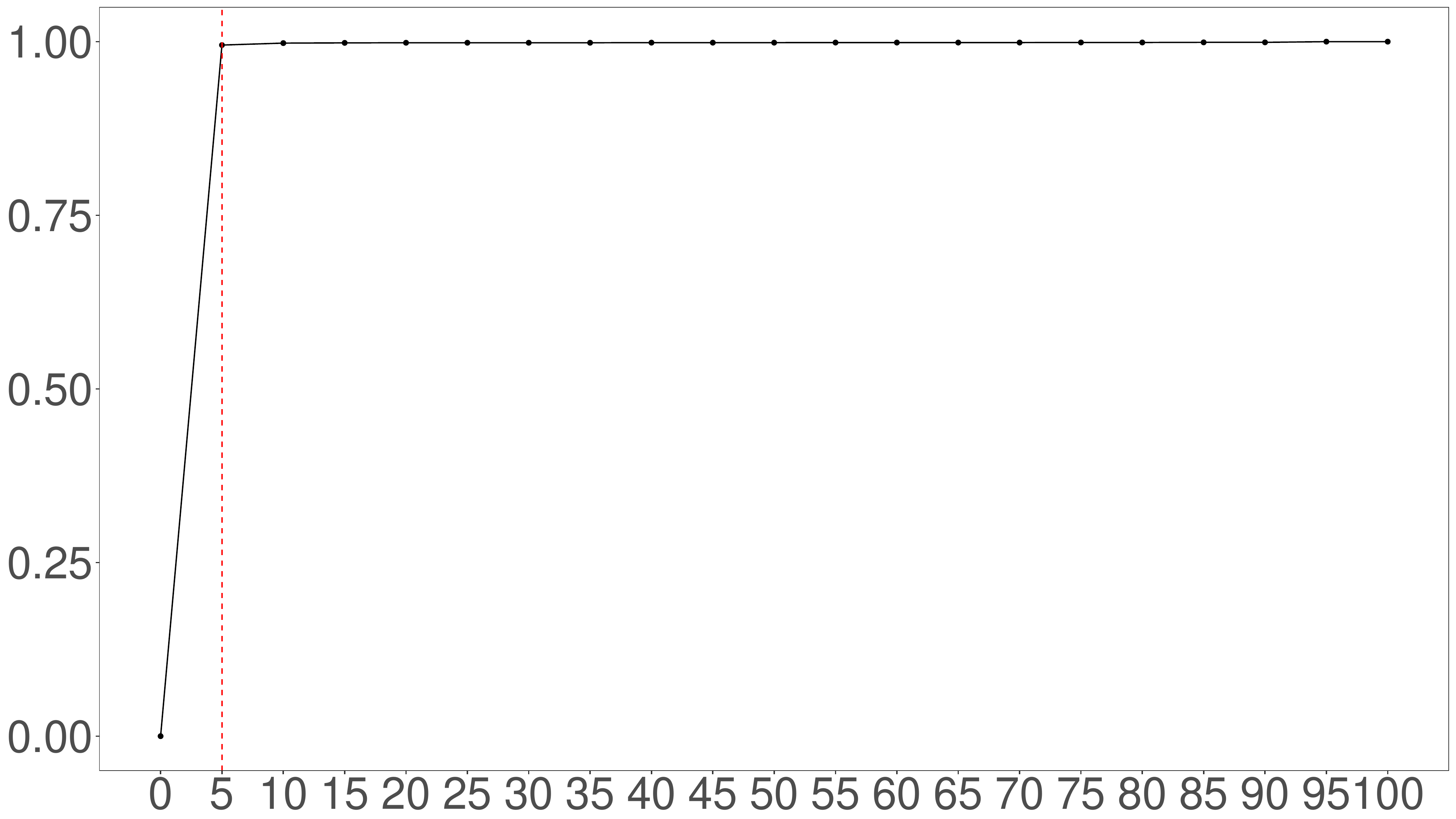}
}\\
\subfloat[HPC]{
    \includegraphics[trim=0 10 0 0,clip,width=.32\columnwidth]{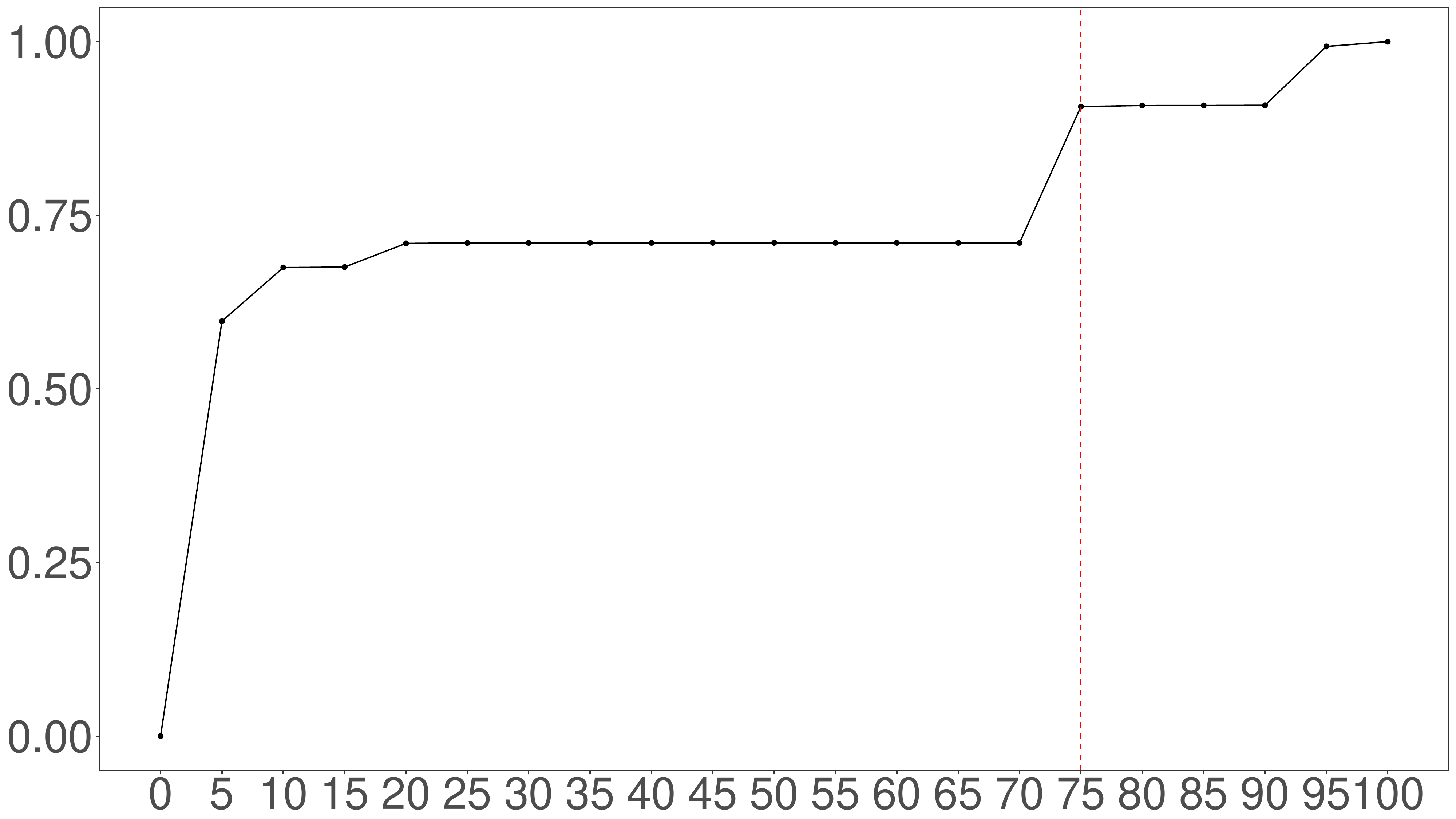}
}
\subfloat[Linux]{
    \includegraphics[trim=0 10 0 0,clip,width=.32\columnwidth]{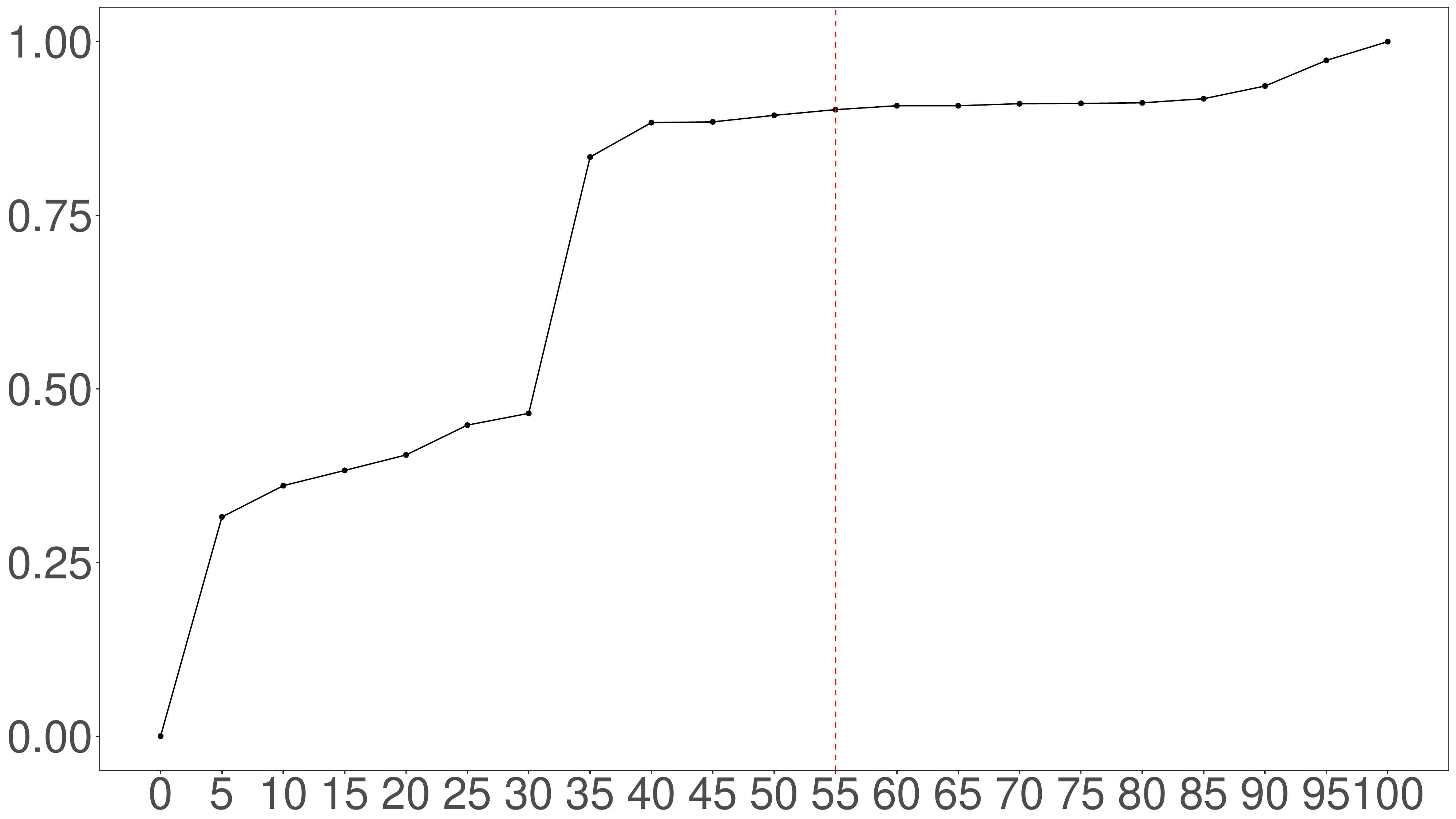}
}
\subfloat[Mac]{
    \includegraphics[trim=0 10 0 0,clip,width=.32\columnwidth]{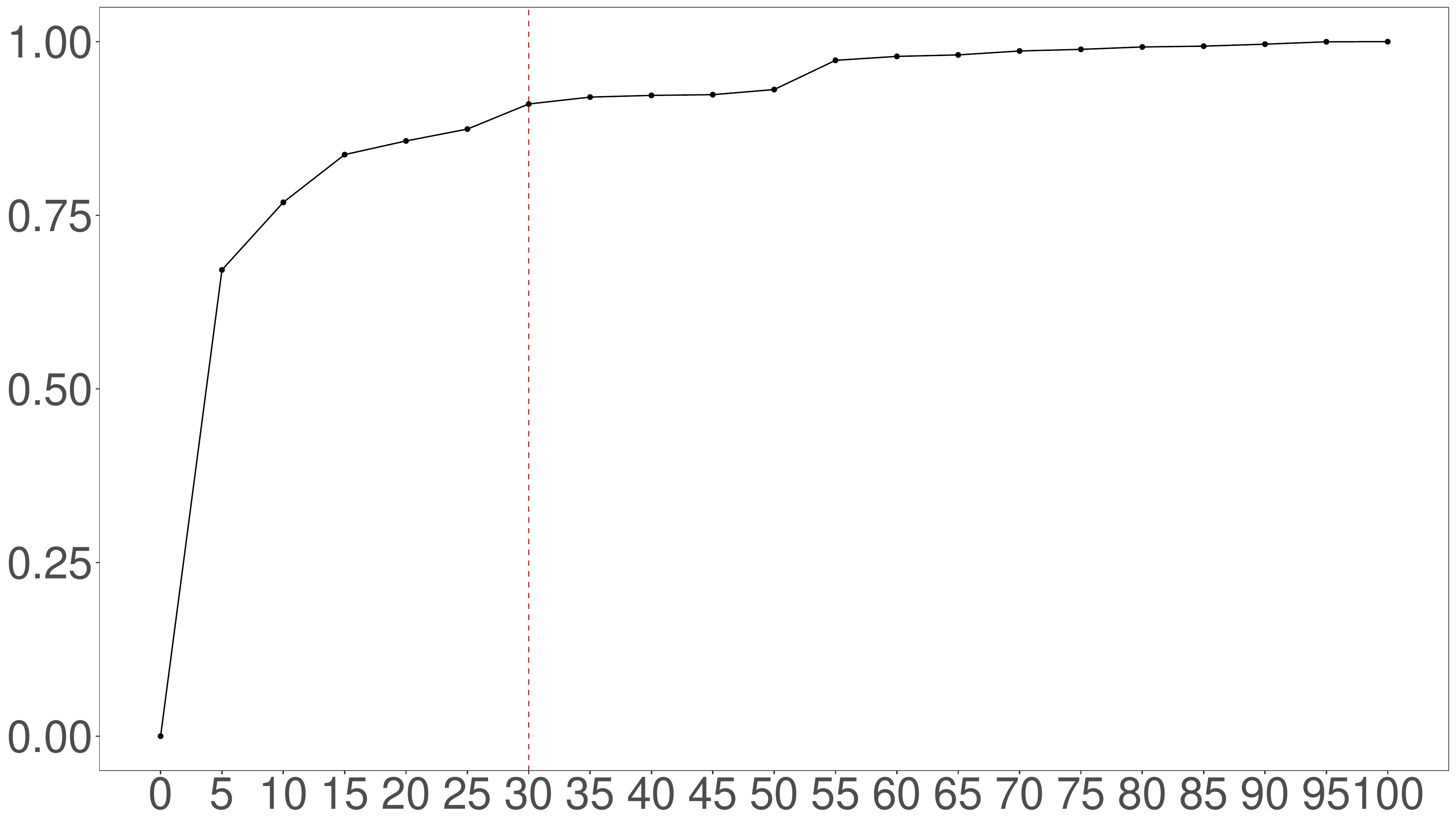}
}\\
\subfloat[OpenSSH]{
    \includegraphics[trim=0 10 0 0,clip,width=.32\columnwidth]{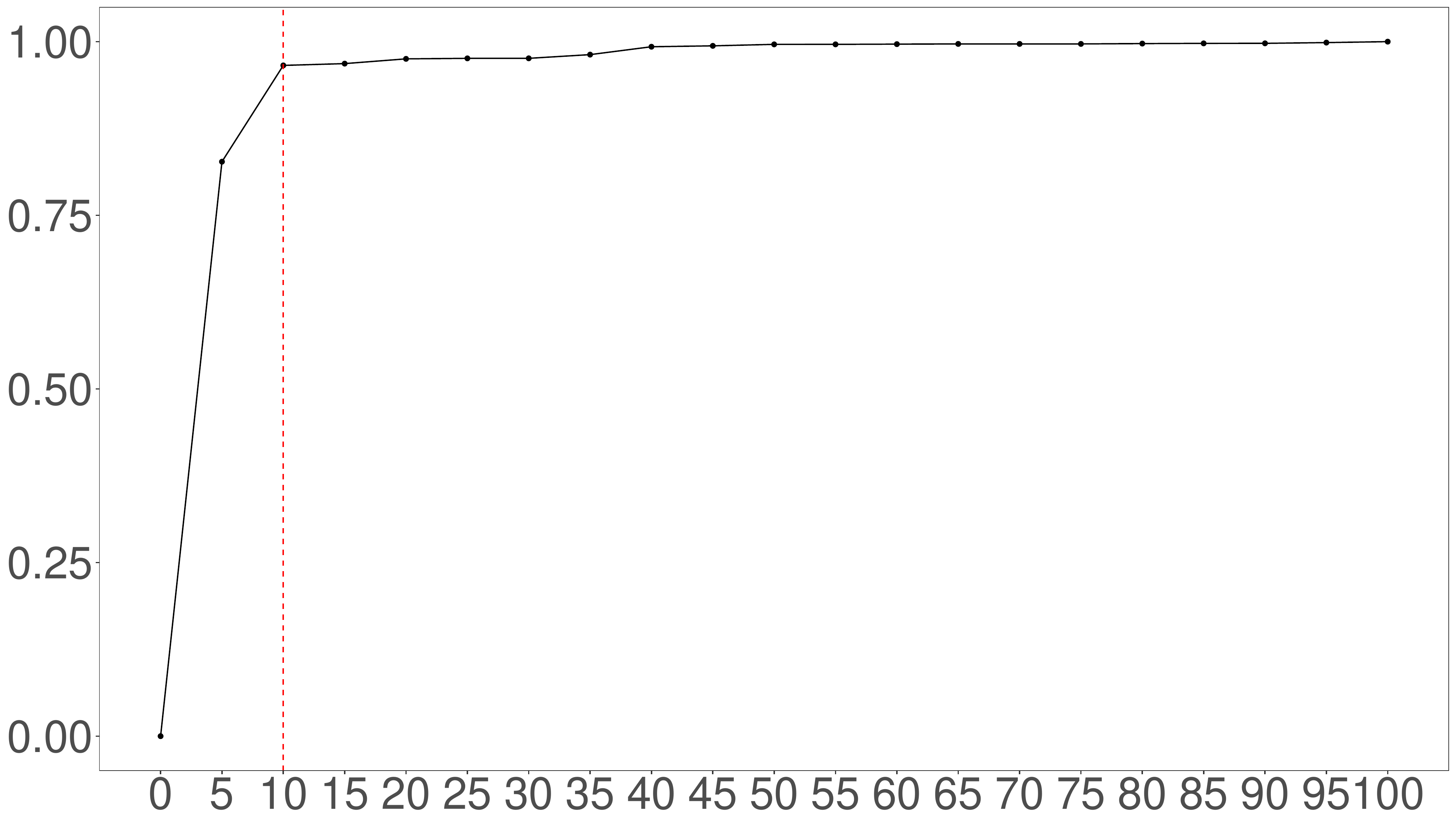}
}
\subfloat[OpenStack]{
    \includegraphics[trim=0 10 0 0,clip,width=.32\columnwidth]{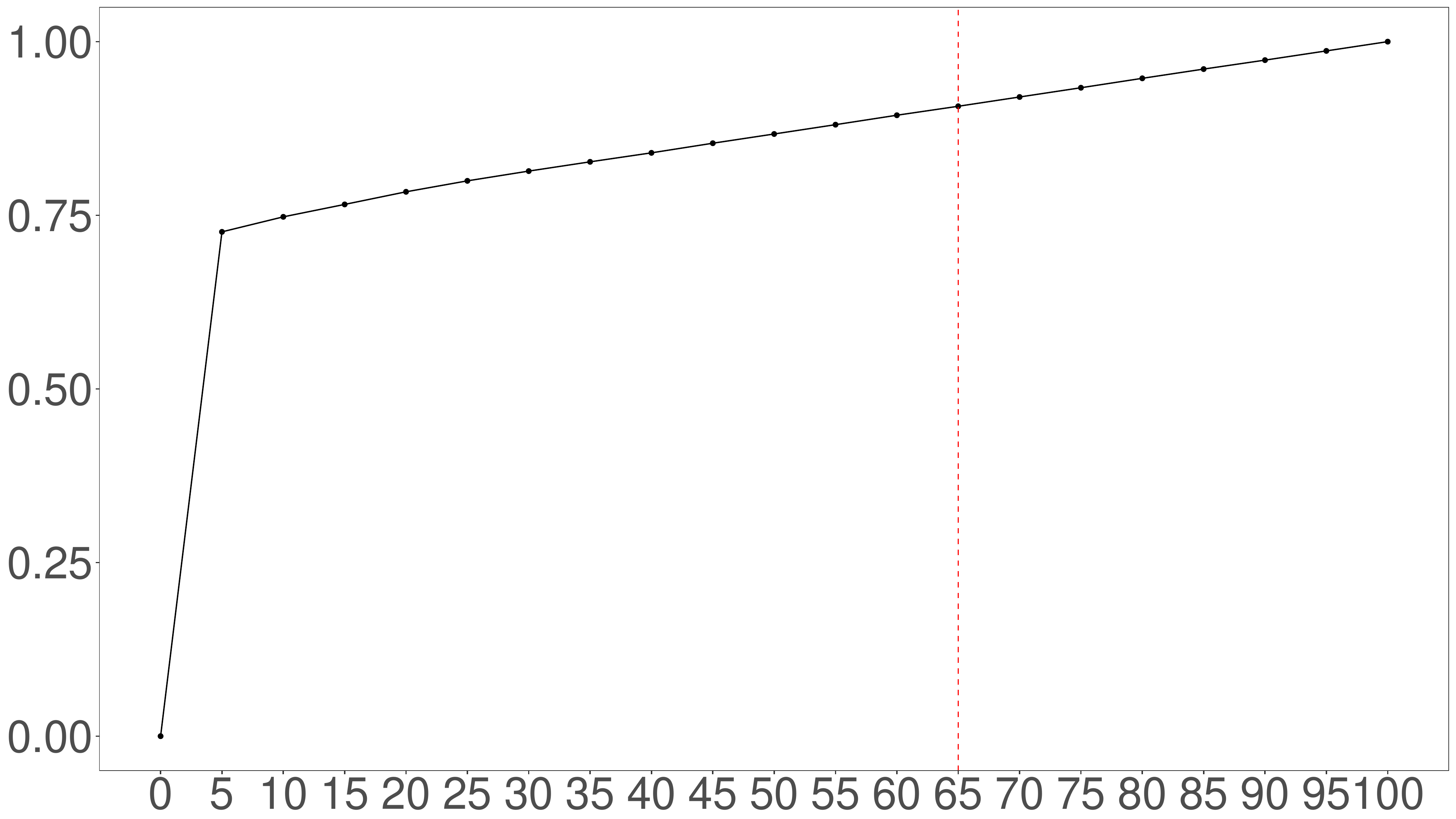}
}
\subfloat[Proxifier]{
    \includegraphics[trim=0 10 0 0,clip,width=.32\columnwidth]{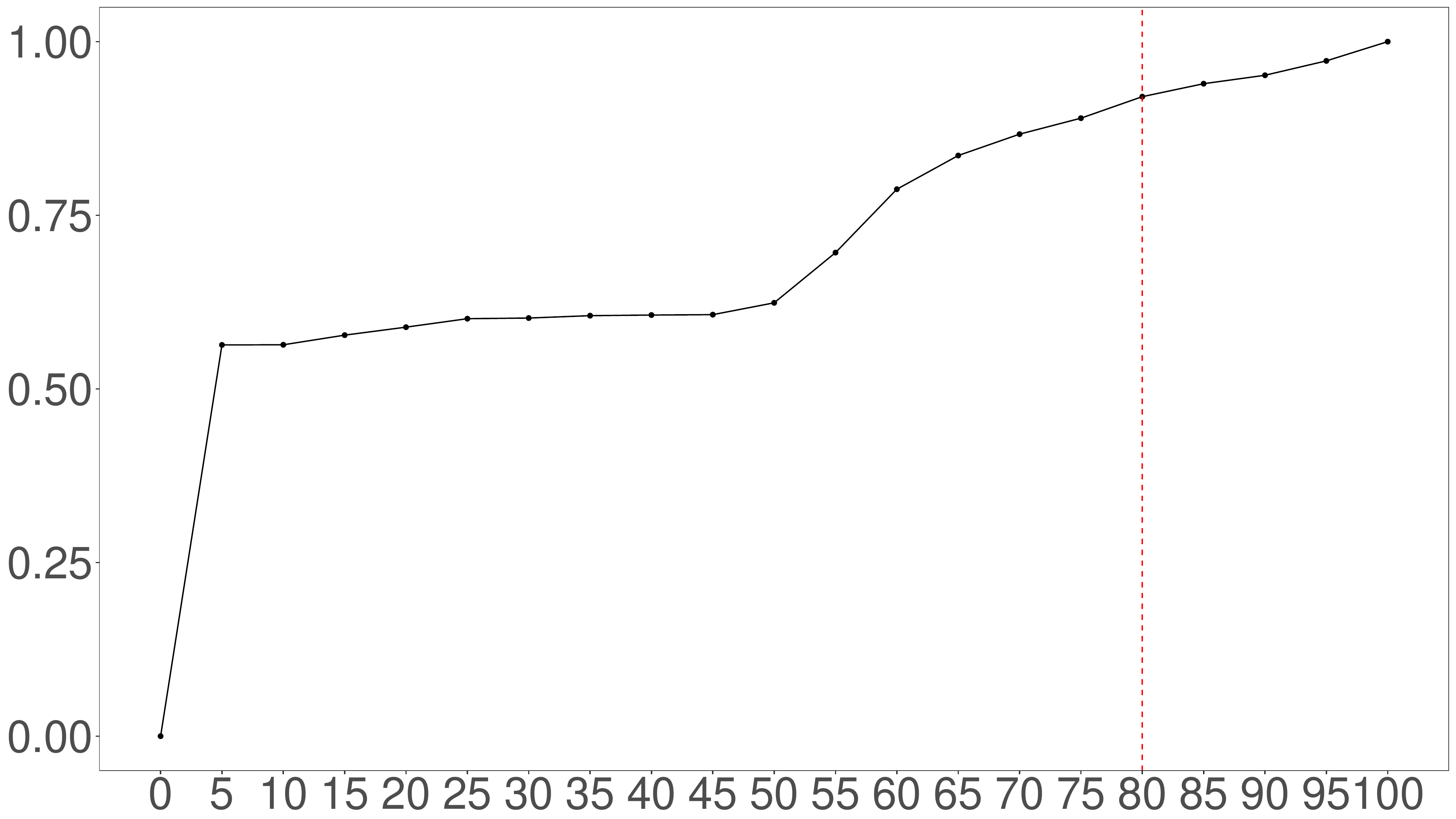}
}\\
\subfloat[Spark]{
    \includegraphics[trim=0 10 0 0,clip,width=.32\columnwidth]{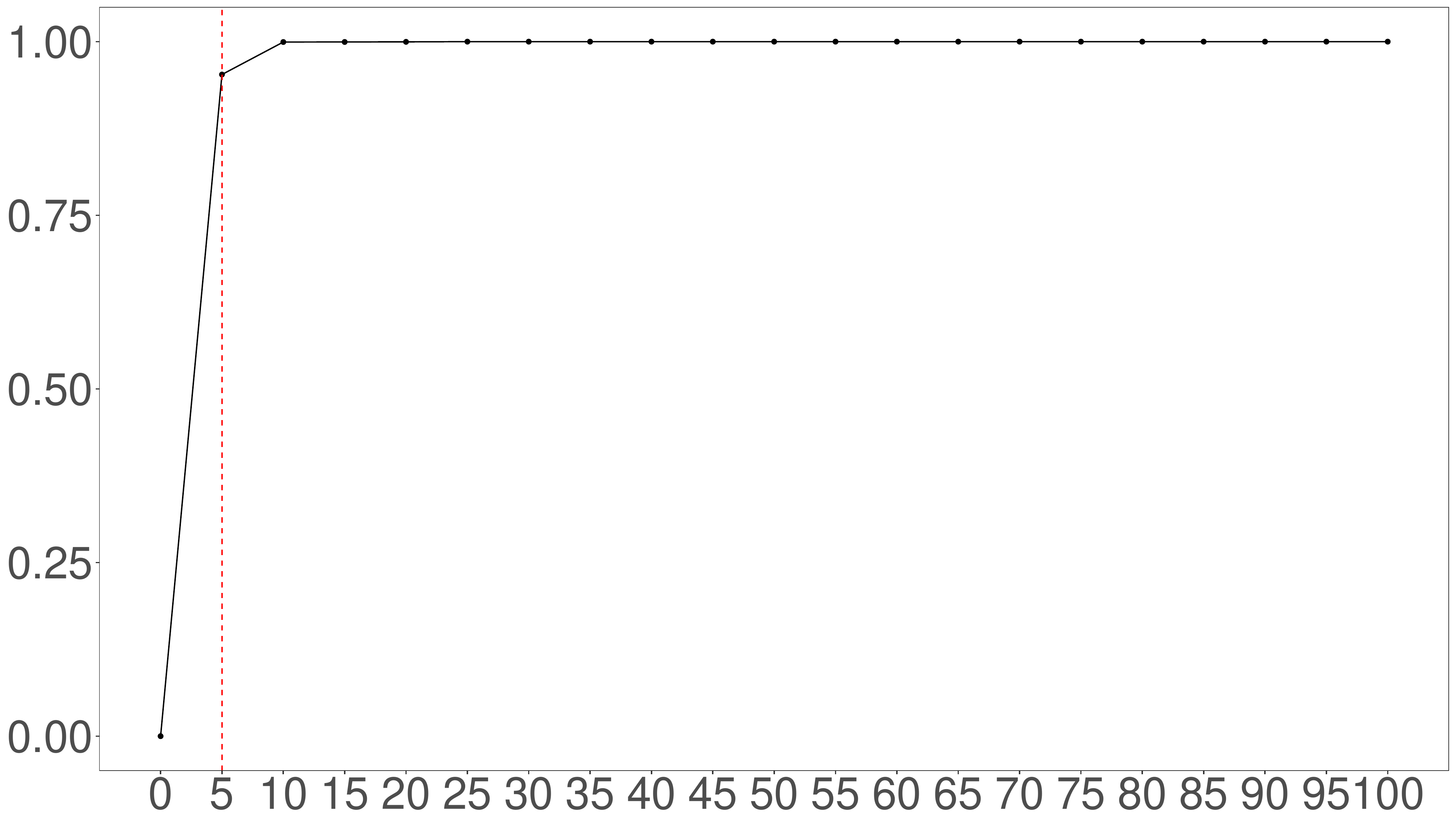}
}
\subfloat[Zookeeper]{
    \includegraphics[trim=0 10 0 0,clip,width=.32\columnwidth]{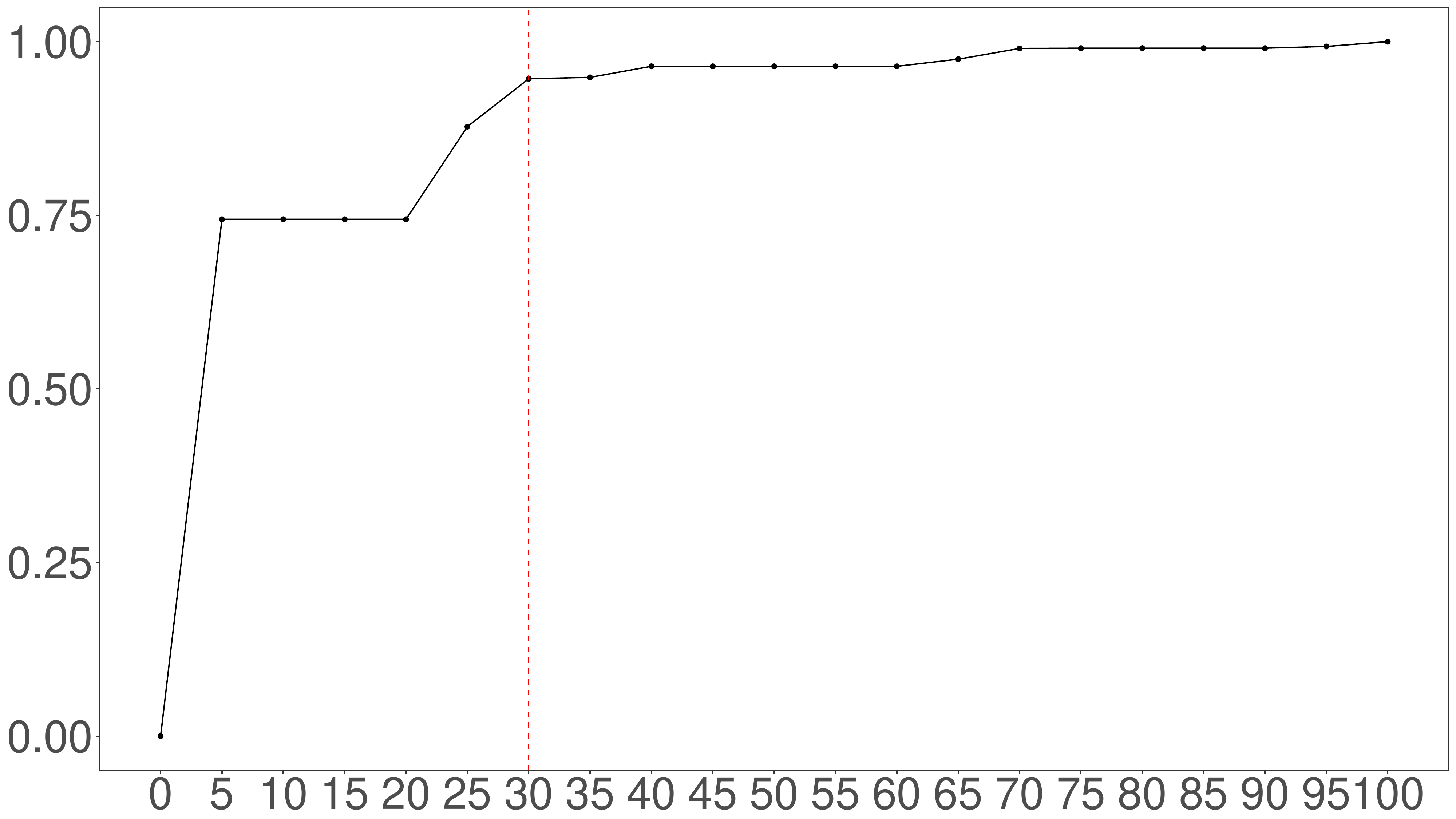}
}
\caption{The agreement ratio of log parsing results with having a part of log to generate dictionary and using all logs to generate dictionary. The red vertical lines indicate that the agreement ratios reach 90\%.}
\label{figure:stable}
\end{figure}

\noindent \textbf{\approach's parsing results are stable with a dictionary generated from a small portion of log data. }
Figure~\ref{figure:stable} shows agreement ratio between parsing results from using partial log data to generate an $n$-gram dictionary and using all log data. The red line in the figures indicates that the agreement ratio is over 90\%. In nine out of 14 studied log data sets, our log parser can generate an $n$-gram dictionary from less than 30\% of the entire log data, while having over 90\% of the log parsing results the same as using all the logs to generate a dictionary. In particular, one of the large log dataset from Spark gains 95.5\% agreement ratio with only first 5\% of the log data. On the one hand, our results show that the log data is indeed repetitive. Such results demonstrate that practitioners can consider leveraging the two parts of \approach in separate, i.e., generating the $n$-gram dictionary (i.e., Figure~\ref{fig:step1}) may not be needed for every log message, while the parsing of each log message (i.e., Figure~\ref{fig:step2}) can depend on a dictionary generated from existing logs. 

We manually check the other five log datasets and we find that in all these data sets, some parts of the log datasets have drastically different log events than others. For example, between the first 30\% and 35\% of the log data in Linux, a large number of log messages are associated with new events for Bluetooth connections and memory issues. Such events do not exist in the logs in the beginning of the dataset. The unseen logs causes parsing results using dictionary from the beginning of the log data to be less agreed with the parsing results using the the entire logs. However, it is interesting to see that after our dictionary learns the $n$-grams in that period, the log parsing results become stable. Therefore, in practice, developers may need to monitor the parsing results to indicate of the need of updating the $n$-gram dictionary from logs.



\subsection{Scalability}
\label{sec:scability}

In order to achieve a high-scalability of log parsing, we migrate \approach to \emph{Spark}. \emph{Spark}~\cite{Zaharia:2010:SCC:1863103.1863113} is an open-source distributed data processing engine, with high-level API in several program languages such as Java, Scala, Python, and R. \emph{Spark} has been adopted widely in practice for analyzing large-scale data including log analysis. We migrate each step of \approach, i.e., 1) generating $n$-gram model based dictionaries and 2) parsing log messages using dictionaries, separately to \emph{Spark}. In particular, the first step of generating dictionary is written similar as a typically \emph{wordcount} example program, where each item in the dictionary is a $n$-gram from a log message. In addition, the second step of parsing log messages is trivial to run in parallel where each log message is parsed independently\footnote{Due to the limited space, the detail of our implementation of the Spark based \approach is available in our replication package.}.

We evaluate the scalability of \approach on a clustering with one master node and five worker nodes running \emph{Spark} 2.43. Each node is deployed on a desktop machine with the same specifications as used in our efficiency evaluation (cf. Section~\ref{sec:efficiency}). In total, our cluster has four cores for each worker, leading to a total of 20 cores for processing logs. We first store the log data into HDFS that are deployed on the cluster with a default option of three replications. We then run the \emph{Spark} based \approach on the \emph{Spark} cluster with one worker (four cores) to five workers (20 cores) enabled. We evaluate the scalability on the same log datasets used for evaluating the efficiency, except for Android due to its relatively small size. We measure the throughput for parsing each log data to assess scalability. Due to the possible noise in a local network environment and the indeterministic nature of the parallel processing framework, we independently repeat each run 10 times when measuring throughput, i.e., number of log messages parsed per second.


\begin{figure}[tbh]
\centering
\subfloat[BGL]{
    \includegraphics[trim=0 5 0 0,clip, width=.49\columnwidth]{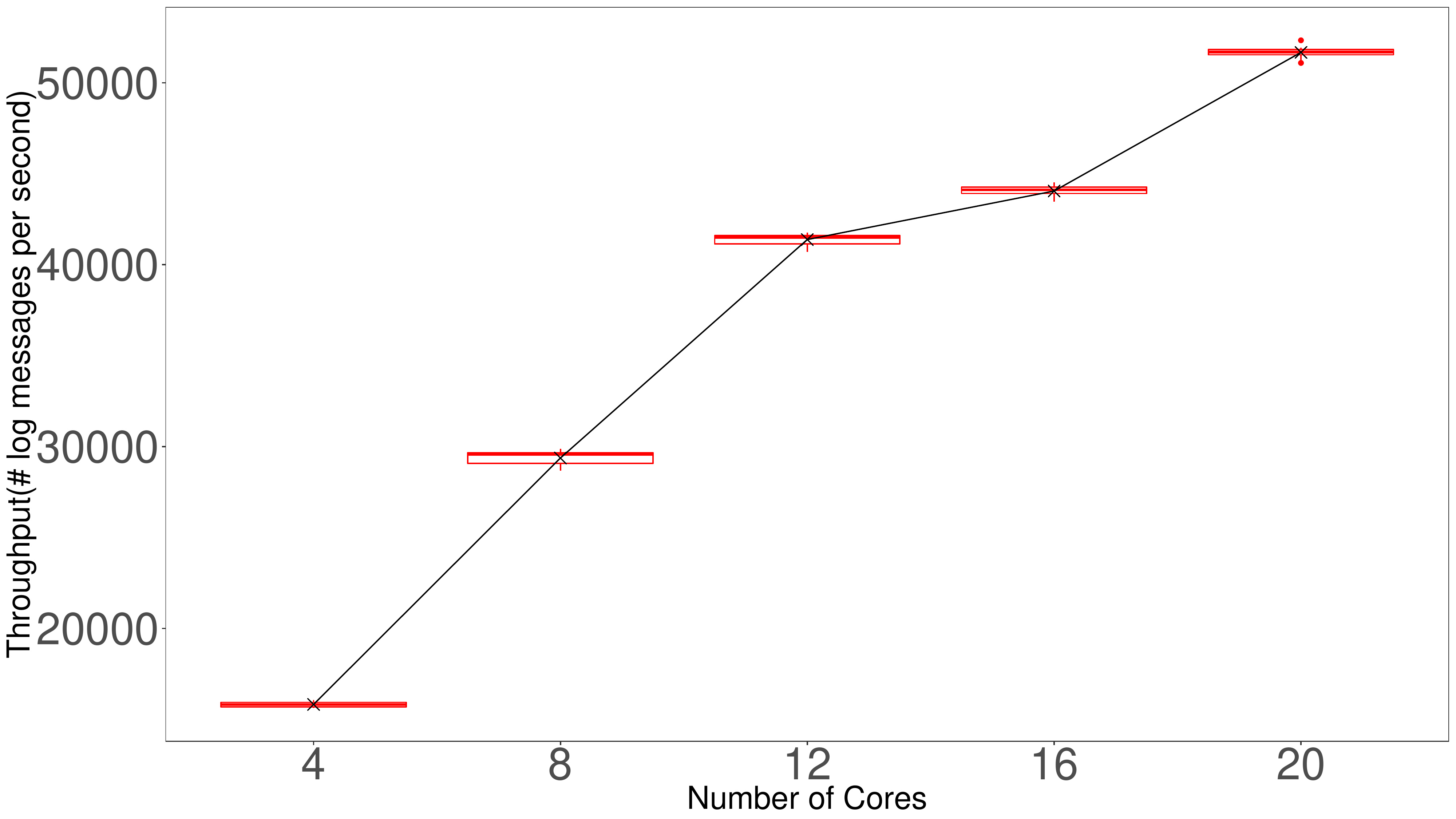}
}
\subfloat[HDFS]{
    \includegraphics[trim=0 5 0 0,clip, width=.49\columnwidth]{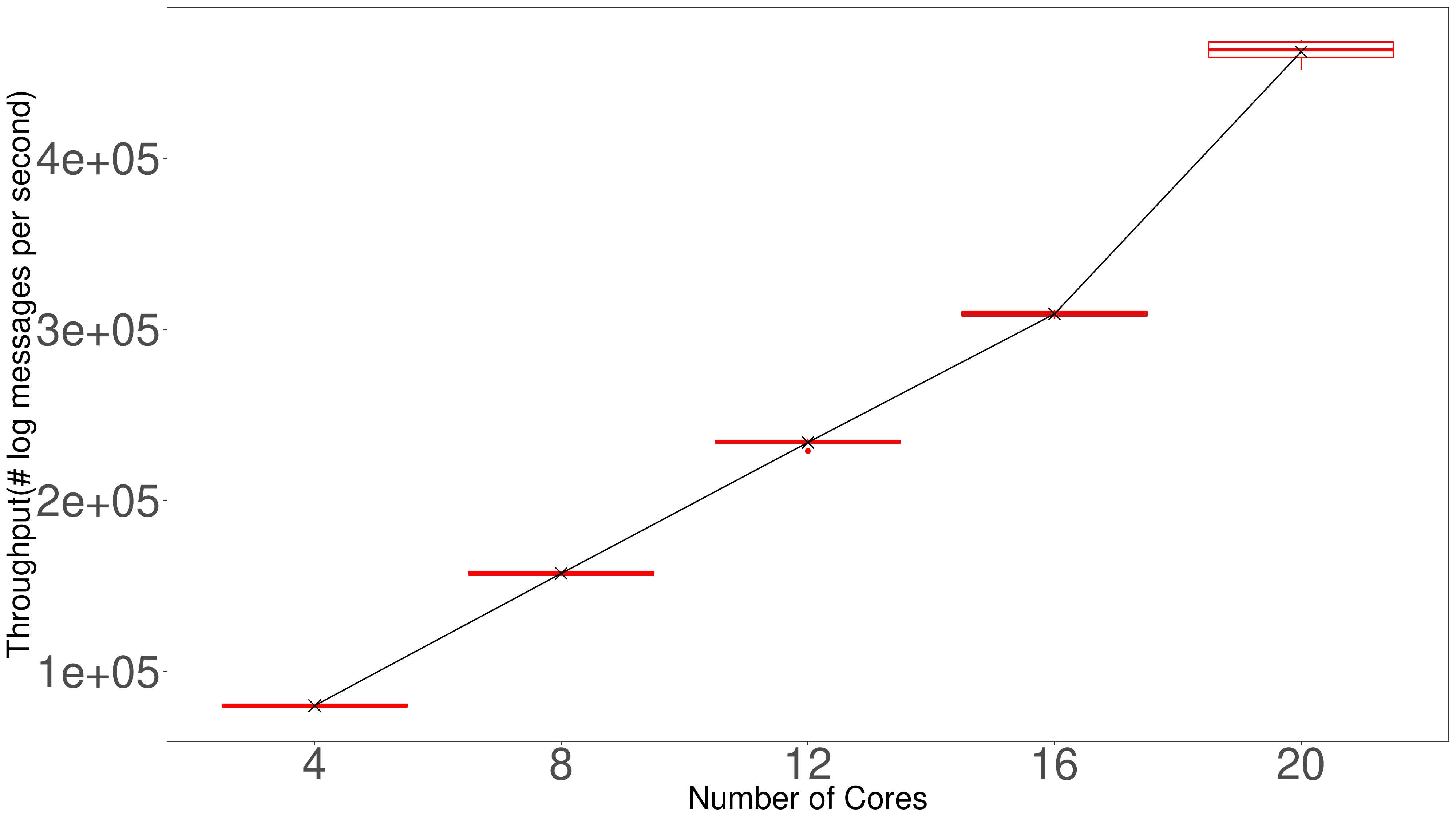}
}\\
\subfloat[Windows]{
    \includegraphics[trim=0 5 0 0,clip, width=.49\columnwidth]{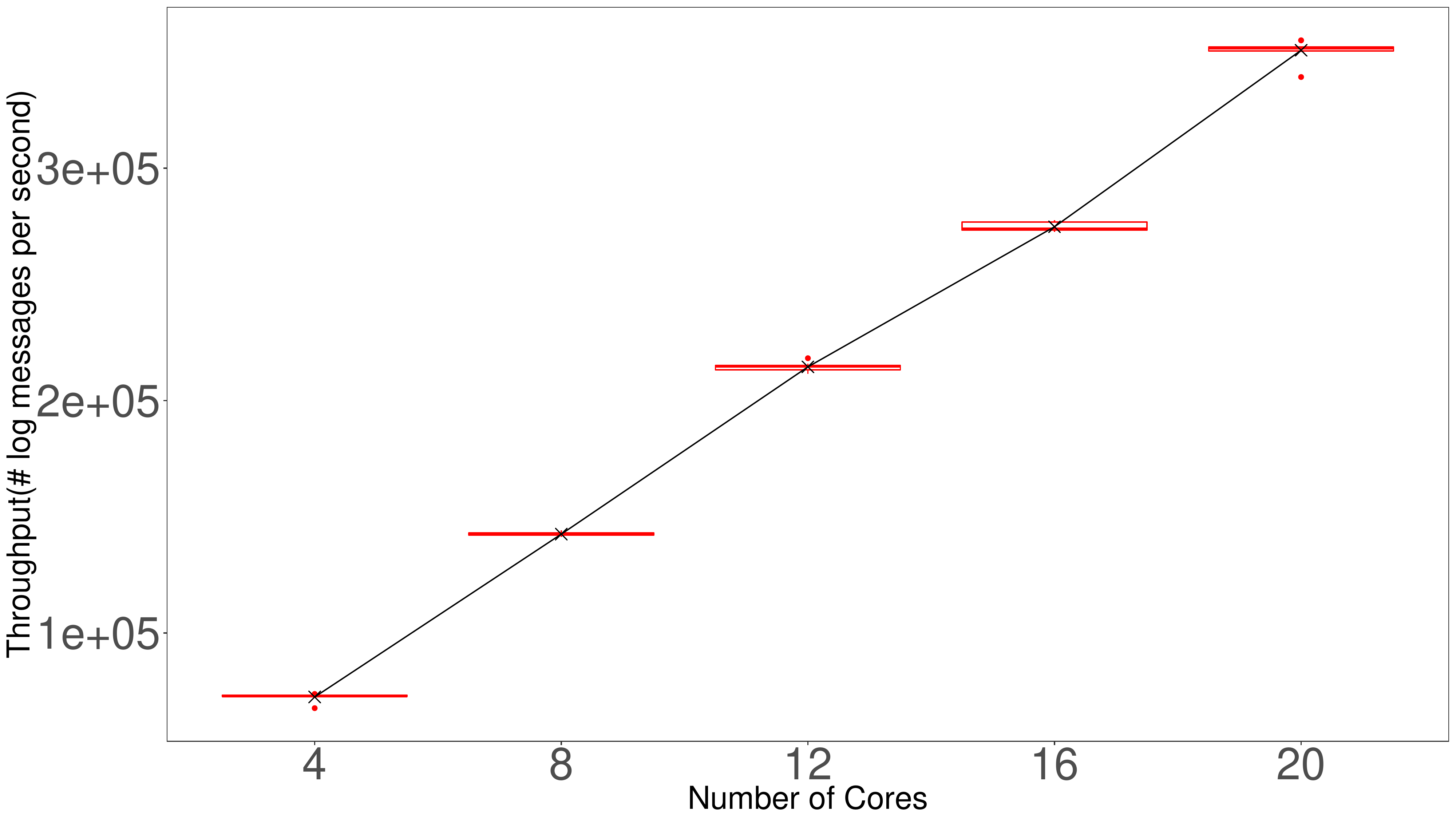}
}
\subfloat[Spark]{
    \includegraphics[trim=0 5 0 0,clip, width=.49\columnwidth]{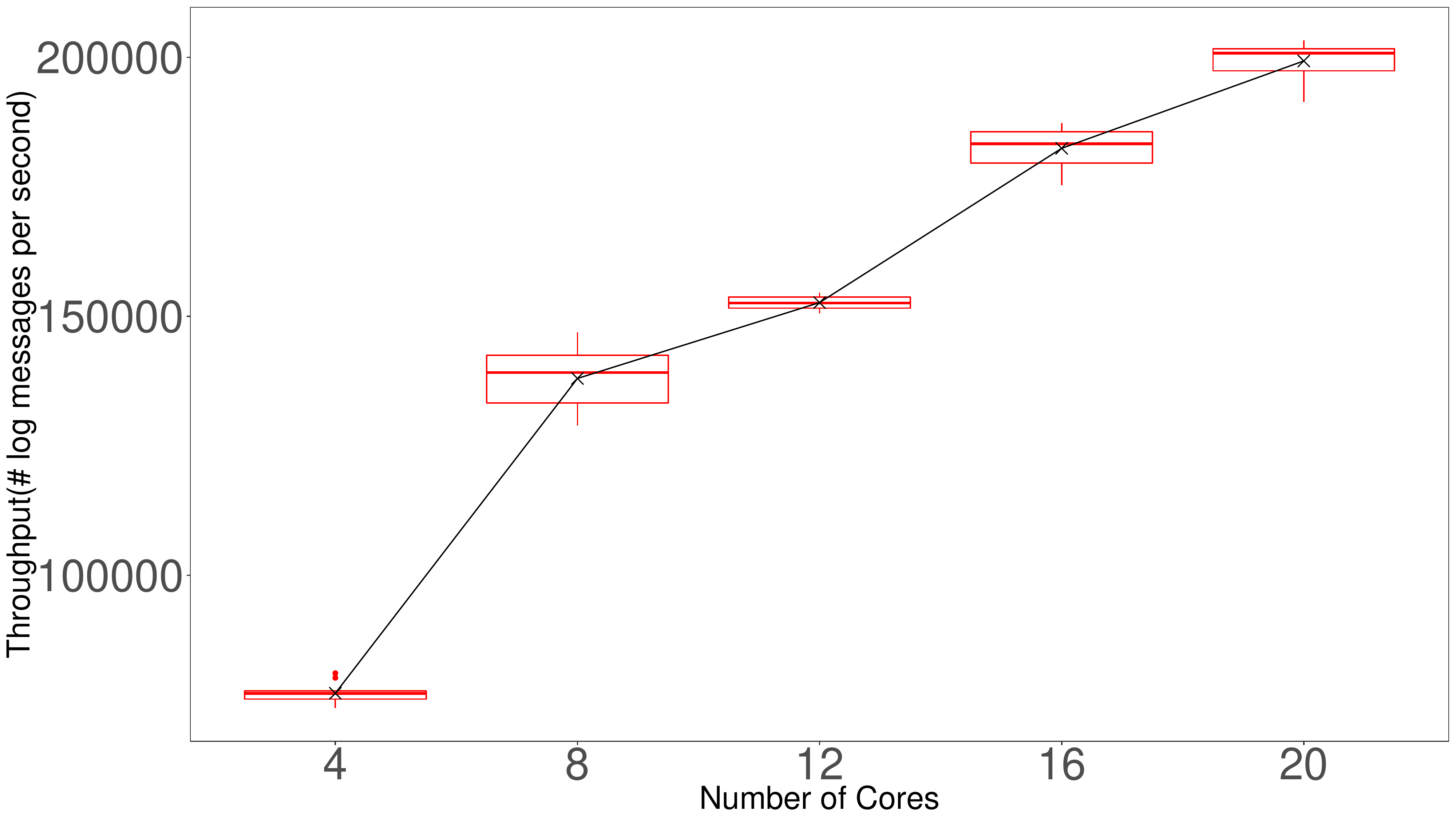}
}
\caption{Box plots of running time of \approach with different number of cores.}
\label{fig:Scalibility}
\end{figure}

\subsubsection*{Results} 

\noindent\textbf{\approach scales out efficiently with the number of \emph{Spark} nodes without sacrificing parsing accuracy.}
Figure~\ref{fig:Scalibility} uses boxplots to present the throughput (i.e., number of log messages parsed per second) of \approach when we increase the number of nodes from one (i.e., four cores) to five (i.e., 20 cores).
As shown in Figure~\ref{fig:Scalibility}, the throughput increases nearly linearly, achieving up to 5.7 times speedup as we increase the number of nodes by a factor of five.
In addition, Figure~\ref{fig:Scalibility} shows that the throughput of \approach has low variance when we repeat the parsing of each log dataset 10 times. We would like to note that the parsing accuracy always keeps the same as we increase the number of nodes. 
When the volume of the parsed logs is very large (e.g., the Windows log data), \approach allows practitioners to increase the speed of log parsing efficiently by adding more nodes without sacrificing any accuracy.

\noindent\textbf{\approach achieves near-linear scalability for some logs but less scalabiltiy on other logs.}
A linear scalability means the throughput increases $K$ times when we increase the number of nodes by a factor of $K$, which is usually the best one usually expects to achieve when scaling an application~\cite{DBLP:journals/queue/GuntherPT15, DBLP:journals/corr/abs-1904-08964}.
The throughput of \approach when parsing the HDFS and Windows logs increases by 5.7 to 4.8 times when we increase the number of nodes from one to five, indicating a near-linear or even super-linear scalability.
However, \approach achieves less scalability when parsing the BGL and Spark logs. 
Specifically, the throughput of \approach when parsing the BGL and Spark logs increases 3.3 and 2.7 times when we increase the number of nodes by a factor of five.


\section{Migrating \approach to an Online Parser}
\label{sec:discussion}


\approach parses logs in two steps: 1) generating $n$-gram dictionaries from logs, and 2) using the $n$-gram dictionaries to parse the logs line by line.
Section~\ref{sec:approach} describes an offline implementation of \approach, in which the step for generating the $n$-gram dictionaries is completely done before the step of parsing logs using the $n$-gram dictionaries (even when we evaluate the ease of stabilisation in Section~\ref{sec:stability}). Therefore, the offline implementation requires all the log data used to generate the $n$-gram dictionaries to be available before parsing.
On the contrary, an online parser parses logs line by line, without an offline training step. An online parser is especially helpful in a log-streaming scenario, i.e., to parse incoming logs in a real-time manner.

\approach naturally supports online parsing, as the $n$-gram dictionaries can be updated efficiently when more logs are continuously added (e.g., in log streaming scenarios).
In our online implementation of \approach, we feed logs in a streaming manner (i.e., feeding one log message each time). 
When reading the first log message, the dictionary is empty (i.e., all the $n$-grams have zero occurrence), so \approach parses all the tokens as dynamic variables.
\approach then creates a dictionary using the $n$-grams extracted from the first log message.
After that, when reading each log message that follows, \approach parses the log message using the existing $n$-gram dictionary. Then, \approach updates the existing $n$-gram dictionary on-the-fly using the tokens in the log message. 
In this way, \approach updates the $n$-gram dictionary and parses incoming logs continuously until all the logs are processed.
Similar to Section~\ref{sec:stability}, we measure the ratio of agreement between the parsing results of the online implementation and the offline implementation. For each log message, we only consider the two parsing results agreeing to each other if they are exactly the same.
We also measure the efficiency of \approach when parsing logs in an online manner relative to the offline mode.
Specifically, we measure the efficiency difference ratio, which is calculated as $\frac{T_\textnormal{online} - T_\textnormal{offline}}{T_\textnormal{offline}}$
where $T_\textnormal{online}$ and $T_\textnormal{offline}$ are the time taken by the online \approach and offline \approach to parse the same log data, respectively. 

\subsubsection*{Results} 

\noindent\textbf{The online mode of \approach achieves nearly the same parsing results as the offline \approach.}
Table~\ref{tab:online-offline-comp} compares the parsing results of \approach between the online and offline modes.
We considered the same five large log datasets as the ones used for evaluating the efficiency of \approach (cf. Section~\ref{sec:efficiency}).
The agreement ratio between the online and offline modes of \approach range from 95.0\% to 100.0\%, indicating that the parsing results of the online \approach are almost identical to the parsing results of the offline \approach.

\noindent\textbf{The online mode of \approach reaches a parsing efficiency similar to the offline \approach.}
Table~\ref{tab:online-offline-comp} also compares the efficiency between the online and offline\approach, for the five considered log datasets with sizes varying from 300KB to 1GB. A positive value of the efficiency difference ratio indicates the online mode is slower (i.e., taking longer time), while a negative value indicates the online mode is even faster.
Table~\ref{tab:online-offline-comp} shows that the efficiency difference ratio ranges from -3.0\% to 8.9\%.
Overall, the online mode of \approach is as efficient as the offline model.
In some cases, the online mode is even faster, because the online mode parses logs with smaller incomplete dictionaries -- thus being queried faster -- compared to the full dictionaries used in the offline mode.

In summary, as the online mode of \approach achieves similar parsing results and efficiency compared to the offline mode, \approach can be effectively used in an online parsing scenario. For example, \approach can be used to parse stream logs in a real-time manner.

\begin{table}[t]
  \centering
  \scriptsize
  \caption{Comparing the parsing results of \approach between the online and offline modes.}
  \vspace{-0.25cm}
  \scalebox{0.9}{
  \begin{tabular}{crrrrrrr}
  \hline
    Subject & \multicolumn{6}{c}{Efficiency Difference Ratio}     & \multicolumn{1}{l}{Agreement with} \\
\cmidrule{2-7}     log dataset & \multicolumn{1}{c}{300k} & \multicolumn{1}{c}{1M} & \multicolumn{1}{c}{10M} & \multicolumn{1}{c}{100M} & \multicolumn{1}{c}{500M} & \multicolumn{1}{c}{1G} & \multicolumn{1}{l}{ offline results} \\
    \hline
    HDFS  & 5.9\%  & 0.0\%  & -2.4\% & -1.5\% & -3.0\% & -0.8\% & 100.0\% \\
    Spark & 0.0\%  & -0.3\%  & -0.6\% & 0.3\%  & -3.0\% & -1.1\% & 99.9\% \\
    Windows & 0.0\%  & 0.0\%  & 1.1\%  & -0.0\%  & -0.2\%  & 0.6\%  & 96.8\% \\
    BGL   & 7.1\%  & 6.7\%  & 7.2\%  & 5.9\%  & 7.4\%  & N/A   & 98.7\% \\
    Android & 5.9\%  & 8.9\%  & 6.6\%  & 6.5\%  & N/A   & N/A   & 95.0\% \\
    \hline
    \end{tabular}}
    Note: a positive value means that \approach is loswer with online parsing than offline.
    \vspace{-0.25cm}
  \label{tab:online-offline-comp}%
\end{table}%

\section{Threats to Validity}
\label{sec:threats}
In this section, we discuss the threat to the validity of our paper.

\noindent \textbf{External validity.}
In this work, we evaluate \approach on 16 log datasets from an existing benchmark~\cite{DBLP:conf/icse/ZhuHLHXZL19}. 
\approach achieves a parsing accuracy higher than 0.9 on about half of the datasets.
We cannot ensure that \approach can achieve high accuracy on other log datasets not tested in this work.
Nevertheless, through an evaluation on logs produced by 16 different systems from different domains (e.g., big data applications and operation systems), we show that \approach achieves similar accuracy as the best existing log parsing approaches with a much faster speed.
Future work can improve our approach to achieve high accuracy on more types of log data.

\noindent \textbf{Internal validity.}
\approach leverages $n$-grams to parse log data.
$n$-grams are typically used to model natural languages or source code that are written by humans. However, logs are different from natural languages or source code as logs are produced by machines and logs contain static and dynamic information.
Nevertheless, we show that $n$-grams can help us effectively distinguish static and dynamic information in log parsing. 
Future work may use $n$-grams to model log messages in other log-related analysis. We use an automated approach to determine the threshold for identifying statically and dynamically generated tokens. Such automatically generated thresholds may not be optimal, i.e., by further optimizing the thresholds, our approach may achieve even higher accuracy; while our currently reported accuracy may not be the highest that our approach can achieve.

\noindent \textbf{Construct validity.}
In the evaluation of this work, we compare \approach with six other log parsing approaches. 
There exists other log parsing approaches (e.g., \emph{LKE}~\cite{DBLP:conf/icdm/FuLWL09}) that are not evaluated in this work.
We only consider five existing approaches as we need to manually verify the parsing accuracy of each approach which takes significant human efforts.
Besides, the purpose of the work is not to provide a benchmark, but rather to propose and evaluate an innovative and promising log parsing approach.
Nevertheless, we compare \approach with the best-performing log parsing approaches evaluated in a recent benchmark~\cite{DBLP:conf/icse/ZhuHLHXZL19}.
Our results show that \approach achieves better parsing accuracy and much faster parsing speed compared to existing state-of-the-art approaches. 

\section{Conclusion}
\label{sec:conclusion}


In this work, we propose \approach, an automated log parsing approach that leverages $n$-grams dictionaries to parse log data in an efficient manner. 
The nature of the $n$-gram dictionaries also enables one to construct the dictionaries in parallel without sacrificing any parsing accuracy and update the dictionaries online when more logs are added (e.g., in log streaming scenarios). 
Through an evaluation of \approach on 16 public log datasets, we demonstrated that \approach can achieve high accuracy and efficiency while parsing logs in a stable and scalable manner. In particular, \approach outperforms the state-of-the-art log parsing approaches in efficiency and achieves better parsing accuracy than existing approaches.
Finally, We demonstrate that \approach can effectively supports online parsing when logs are continuously generated as a stream with similar parsing results and efficiency to the offline mode.
This is the fist work that uses $n$-grams in log analysis, which demonstrates a success on leveraging a mix of the (un)natural characteristics of logs. 
\approach can benefit future research and practices that rely on automated log parsing to achieve their log analysis goals.

\bibliographystyle{IEEEtran}
\bibliography{bibliography}

\begin{IEEEbiography}[{\includegraphics[width=1in,height=1.25in,clip,keepaspectratio]{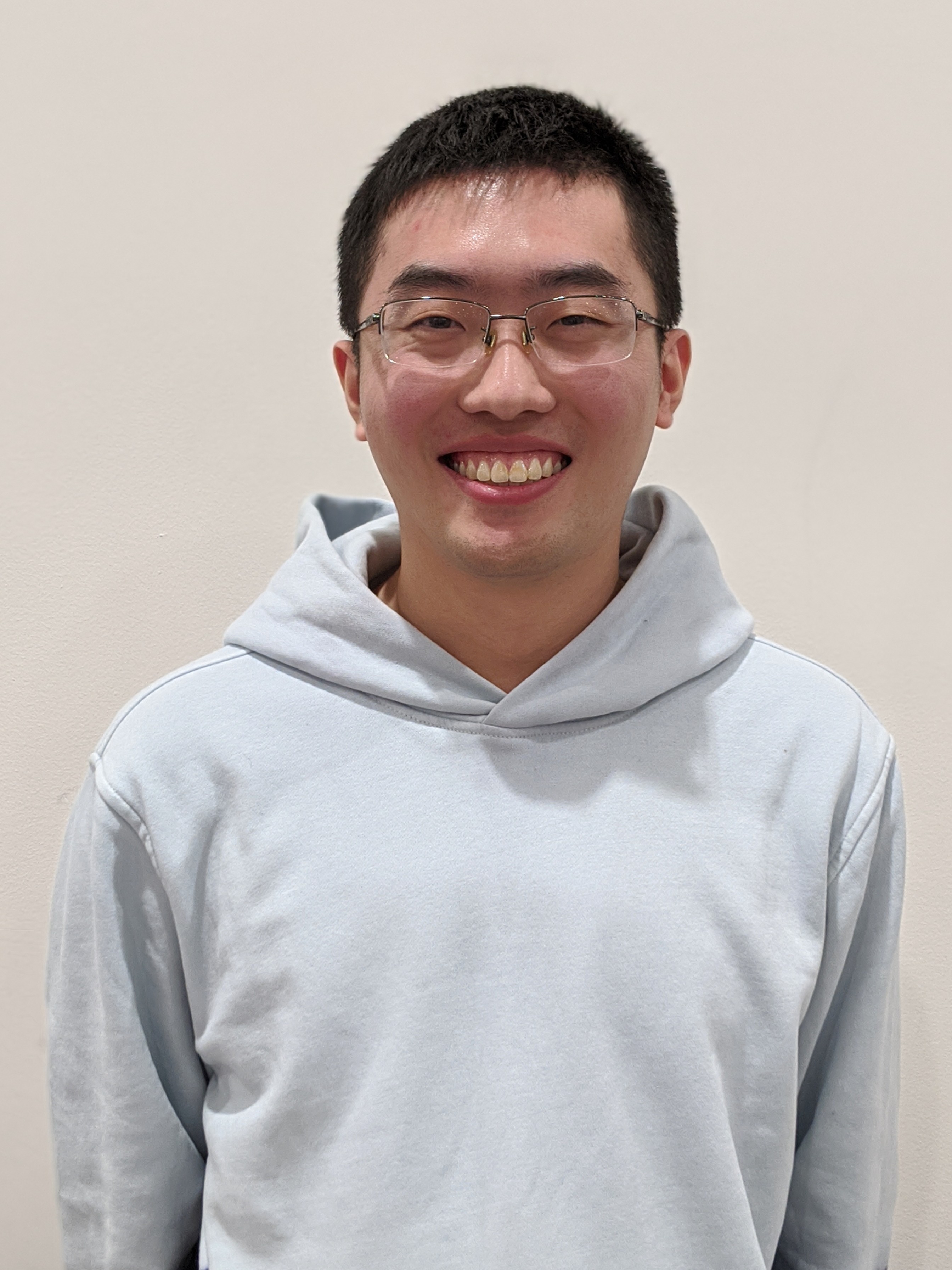}}]{Hetong Dai}
is a Master student in the Department of Computer Science and Software Engineering at Concordia University, Montreal, Canada, supervised by Weiyi Shang. His research lies within Software Engineering, with special interests in software engineering for ultra-large-scale systems, software log mining and mining software repositories. He obtained his BS from Nanjing University. Contact him at he\_da@encs.concordia.ca
\end{IEEEbiography}

\begin{IEEEbiography}[{\includegraphics[width=1in,height=1.25in,clip,keepaspectratio]{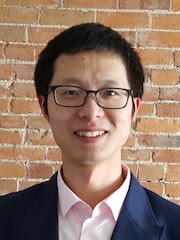}}]{Heng Li}
is a postdoctoral fellow in the School of Computing at Queen's University, Canada. His research lies within Software Engineering and Computer Systems, with special interests in Artificial Intelligence for DevOps, software log mining, software performance engineering, mining software repositories, and qualitative studies of software engineering data. He obtained his BE from Sun Yat-sen University, MSc from Fudan University, and PhD from Queen's University, Canada. He worked at Synopsys as a full-time R\&D Engineer before starting his PhD. Contact him at hengli@cs.queensu.ca
\end{IEEEbiography}

\begin{IEEEbiography}[{\includegraphics[width=1in,height=1.25in,clip,keepaspectratio]{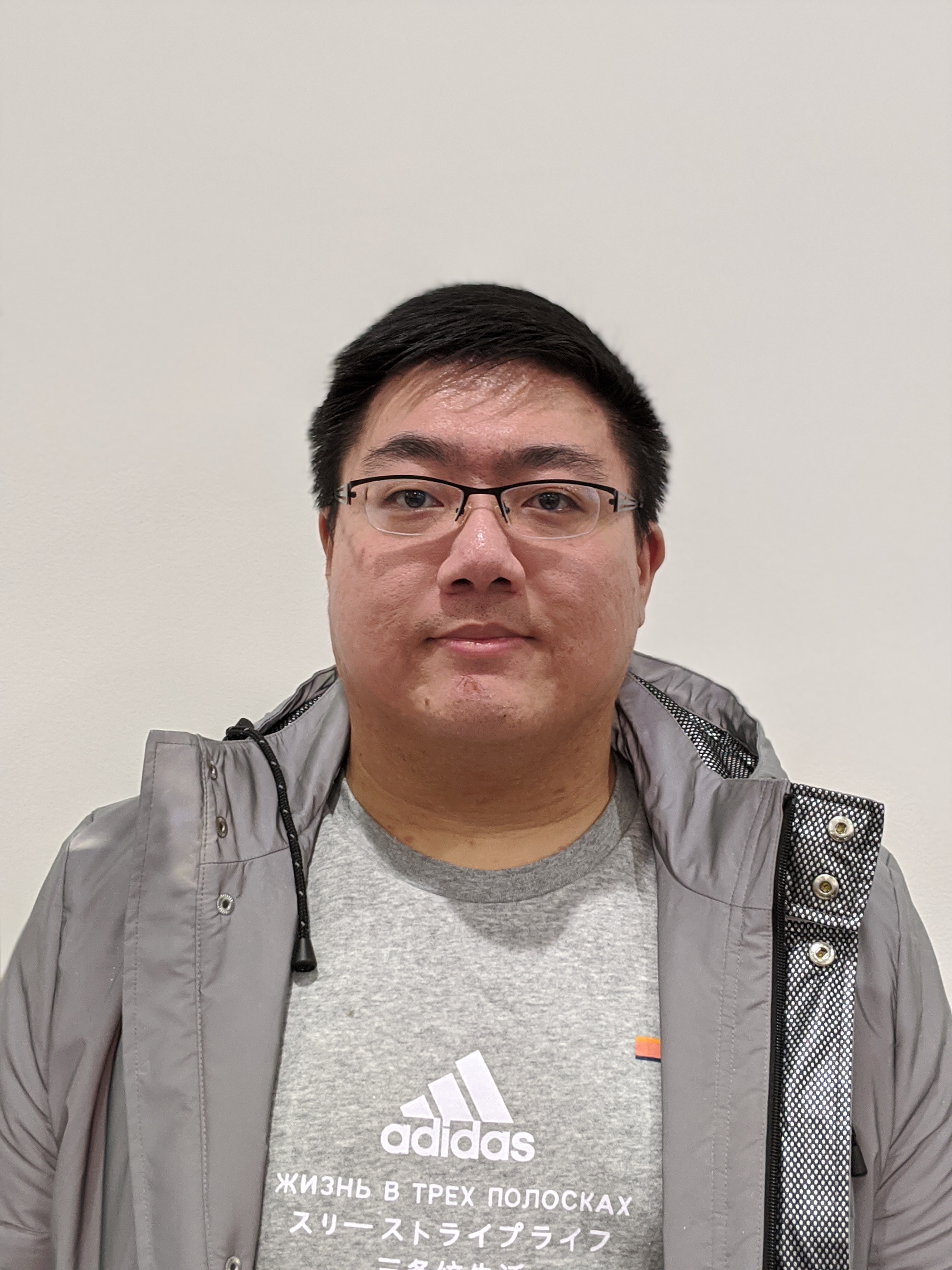}}]{Che Shao Chen}
is studying Master of Software Engineering at Concordia University, Montreal, Canada. His interests and research areas are Software Engineering with special interests in software engineering for ultra-large-scale systems related to Software Refactoring, Data Mining, software log mining, and mining software repositories. He is supervised by Professor Weiyi Shang, who is an assistant professor and research chair at Concordia University. He obtained his BS from Tamkang University. Contact him at c\_chesha@encs.concordia.ca
\end{IEEEbiography}

\begin{IEEEbiography}[{\includegraphics[width=1in,height=1.25in,clip,keepaspectratio]{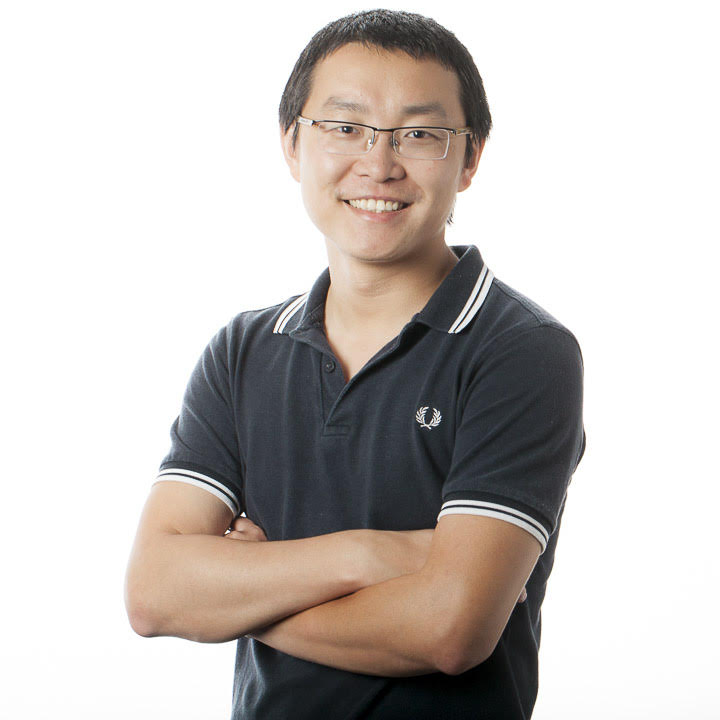}}]{Weiyi Shang}
is an Assistant Professor and Concordia University Research Chair in Ultra-large-scale Systems at the Department of Computer Science and Software Engineering at Concordia University, Montreal. He has received his Ph.D. and M.Sc. degrees from Queen’s University (Canada) and he obtained B.Eng. from Harbin Institute of Technology. His research interests include big data software engineering, software engineering for ultra-large–scale systems, software log mining, empirical software engineering, and software performance engineering. His work has been published at premier venues such as ICSE, FSE, ASE, ICSME, MSR and WCRE, as well as in major journals such as TSE, EMSE, JSS, JSEP and SCP. His work has won premium awards, such as SIGSOFT Distinguished paper award at ICSE 2013 and best paper award at WCRE 2011. His industrial experience includes helping improve the quality and performance of ultra-large-scale systems in BlackBerry. Early tools and techniques developed by him are already integrated into products used by millions of users worldwide. Contact him at shang@encs.concordia.ca; http://users.encs.concordia.ca/∼shang.
\end{IEEEbiography}

\begin{IEEEbiography}[{\includegraphics[width=1in,height=1.25in,clip,keepaspectratio]{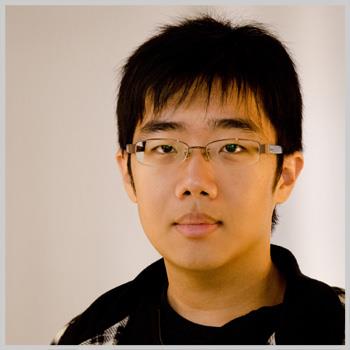}}]{Tse-Hsun (Peter) Chen}
is an Assistant Professor in the Department of Computer Science and Software Engineering at Concordia University, Montreal, Canada. He leads the Software PErformance, Analysis, and Reliability (SPEAR) Lab, which focuses on conducting research on performance engineering, program analysis, log analysis, production debugging, and mining software repositories. His work has been published in flagship conferences and journals such as ICSE, FSE, TSE, EMSE, and MSR. He serves regularly as a program committee member of international conferences in the field of software engineering, such as ASE, ICSME, SANER, and ICPC, and he is a regular reviewer for software engineering journals such as JSS, EMSE, and TSE. Dr. Chen obtained his BSc from the University of British Columbia, and MSc and PhD from Queen's University. Besides his academic career, Dr. Chen also worked as a software performance engineer at BlackBerry for over four years. Early tools developed by Dr. Chen were integrated into industrial practice for ensuring the quality of large-scale enterprise systems. More information at: http://petertsehsun.github.io/.
\end{IEEEbiography}

\end{document}